\documentclass[aps,prd,reprint,twocolumn,floatfix,preprintnumbers]{revtex4-2}
\usepackage{graphicx,color,xcolor}
\usepackage{amsmath,amssymb,amsfonts}
\usepackage[caption=false]{subfig}
\usepackage[colorlinks=True, citecolor=blue, urlcolor=blue, linkcolor=blue]{hyperref}
\newcommand{\be}{\begin{equation}}
\newcommand{\ee}{\end{equation}}
\newcommand{\ba}{\begin{eqnarray}}
\newcommand{\ea}{\end{eqnarray}}
\newcommand{\nn}{\nonumber}
\newcommand{\Tr}{\mathrm{Tr}}
\newcommand{\jmax}{j_\mathrm{max}}

\begin{document}
\title{Eigenstate Thermalization in 2+1 dimensional SU(2) Lattice Gauge Theory}
\author{Lukas Ebner}
\affiliation{Institut f\"ur Theoretische Physik, Universit\"at Regensburg, D-93040 Regensburg, Germany}
\author{Berndt M\"uller}
\affiliation{Department of Physics, Duke University, Durham, North Carolina 27708, USA}
\author{Andreas Sch\"afer}
\affiliation{Institut f\"ur Theoretische Physik, Universit\"at Regensburg, D-93040 Regensburg, Germany}
\author{Clemens Seidl}
\affiliation{Institut f\"ur Theoretische Physik, Universit\"at Regensburg, D-93040 Regensburg, Germany}
\author{Xiaojun Yao}
\affiliation{InQubator for Quantum Simulation, Department of Physics,
University of Washington, Seattle, Washington 98195, USA}
\date{\today}
\preprint{IQuS@UW-21-062}
\begin{abstract}
We present preliminary numerical evidence for the hypothesis that the Hamiltonian SU(2) gauge theory discretized on a lattice obeys the Eigenstate Thermalization Hypothesis (ETH). To do so we study three approximations: (a) a linear plaquette chain in a reduced Hilbert space limiting the electric field basis to $j=0,\frac{1}{2}$ , (b) a two-dimensional honeycomb lattice with periodic or closed boundary condition and the same Hilbert space constraint, and (c) a chain of only three plaquettes but such a sufficiently large electric field Hilbert space ($j \leq \frac{7}{2})$ that convergence of all energy eigenvalues in the analyzed energy window is observed. While an unconstrained Hilbert space is required to reach the continuum limit of SU(2) gauge theory, numerical resource constraints do not permit us to realize this requirement for all values of the coupling constant and large lattices. In each of the three studied cases we check first for random matrix theory (RMT) behavior in the eigenenergy spectrum and then analyze the diagonal as well as the off-diagonal matrix elements between energy eigenstates for a few operators. Within current uncertainties all results for (a), (b) and (c) agree with ETH predictions. Furthermore, we find the off-diagonal matrix elements of the electric energy operator exhibit RMT behavior in frequency windows that are small enough in (b) and (c). To unambiguously establish ETH behavior and determine for which class of operators it applies, an extension of our investigations is necessary. 
\end{abstract}
\maketitle
\section{Introduction}
\label{sec:Intro}

The question of how thermal behavior emerges in systems governed by the strong interaction has a long history, dating back to Fermi's statistical \cite{Fermi1950high} and Landau's hydrodynamical \cite{Landau:1953gs} models of multiparticle production and ultimately culminating in Hagedorn's statistical bootstrap model \cite{Hagedorn:1965st}. While these models posited microcanonical or canonical thermalization, they did not explain its origins. More recent attempts to show that systems whose dynamics is governed by nonabelian gauge theory thermalize rapidly are based on three approaches: (1) kinetic theory applicable to weak coupling \cite{Kurkela:2011ti,Fu:2021jhl,Brewer:2022vkq}; (2) semiclassical dynamics in holographic duals of strongly coupled supersymmetric gauge theories \cite{Balasubramanian:2010ce,Balasubramanian:2011ur,Waeber:2022vgf}; (3) classical simulations of Hamiltonian SU(2) and SU(3) lattice gauge theories \cite{Muller:1992iw,Gong:1992yu}. All these approaches have their limitations and do not fully answer the question.

The numerically established fact that SU(2) lattice gauge theory is extensively chaotic at the classical level \cite{Bolte:1999th} suggests that the theory also exhibits quantum chaos when quantized. It is generally believed that chaotic quantum systems satisfy the Eigenstate Thermalization Hypothesis (ETH) \cite{Deutsch:1991qu,Srednicki:1994mfb}, which posits that the matrix elements of generic operators in the energy eigenstate basis are given by:
\be
\langle E_\alpha |{\cal A}| E_\beta \rangle = \langle {\cal A} \rangle_{\rm mc} (E)\delta_{\alpha\beta} + e^{-S(E)/2} f_{\cal A}(E,\omega) R_{\alpha\beta} \,,\!
\label{eq:ETH}
\ee
where $E = (E_\alpha + E_\beta)/2$ and $\omega = E_\alpha - E_\beta$, the $R_{\alpha\beta}$ vary radically with zero mean and unit variance, forming a Gaussian distribution in large systems, $\langle {\cal A} \rangle_{\rm mc}(E)$ is the microcanonical expectation value of ${\cal A}$, $f_{\cal A}(E,\omega)$ is the spectral response function of the operator, and the exponential prefactor accounts for the energy level density in terms of the microcanonical entropy $S(E)$ (see \cite{DAlessio:2015qtq,mori2018thermalization} for reviews).

It is often stated (see e.g., \cite{DAlessio:2015qtq}) that ETH behavior is a generalization of random matrix theory (RMT). It is worth noting that the matrix $R_{\alpha\beta}$ may not be a random matrix due to the constraints of the underlying theory. Furthermore, it is important to recognize that RMT expresses statistical properties that may apply to any type of mathematical object, like the zeros of the Riemann zeta function, but the ETH relation Eq.~(\ref{eq:ETH}) expresses a {\it dynamical} property that is encoded in the spectral function $f(E,\omega)$.

While Eq.~(\ref{eq:ETH}) is at the center of a huge body of work, it raises many important questions that are still under debate. Some of these are:
\begin{itemize} 
\item 
For which operators ${\cal A}$ does the defining ETH relation Eq.~(\ref{eq:ETH}) apply? Most likely, this is only the case for certain classes of ``physical'' operators, such as local operators or sufficiently averaged ones. For nonabelian gauge theories a restriction to gauge invariant and multiplicatively renormalizable operators also seems a natural requirement. As such constrained operators first come to mind and thus are typically studied in numerical investigations, it could be that such numerical studies provide only a biased perspective and should not be blindly assumed to generalize. 
\item 
What is the thermalization time for a given operator ${\cal A}$? It was shown that if the ETH is satisfied in the weak sense (i.e., most eigenstates satisfy the ETH), a two-point correlation function factorizes at late time \cite{Alhambra:2019ilc}. Thermalization happens rapidly for most observables \cite{Malabarba:2014yxl}. However, the thermalization time in general depends on the initial state. For some pathological initial states with narrow energy spread, it can be arbitrarily long \cite{Dymarsky:2018sef, Goldstein:2013}. 
\item 
To which precision must Eq.~(\ref{eq:ETH}) be fulfilled to reach a certain degree of thermalization? For example, if equilibrium is reached at late time, only states in a small $\omega$ window are probed.
\end{itemize}
At present we are unable to answer such questions rigorously, making numerical studies of specific systems the most promising avenue to pursue. Comparing the results with phenomenology allows to check the plausibility of the assumptions made and helps to develop a better understanding of the systems studied, without reaching mathematical rigor. Our contribution is of this kind. 

Recently there have been many numerical studies of the ETH using classical digital computers \cite{rigol2008thermalization,PhysRevE.90.052105,mondaini2016eigenstate,Jansen2019ETH,Richter:2020bkf,Mueller:2021gxd}, which are generally limited to rather small systems, e.g., $O(20)$ Ising spins, because the size of the Hilbert space basis generally grows exponentially with the number of degrees of freedom. Even so, these studies provided strong indications that multi-particle states with ergodic dynamics (on the classical level) obey Eq.~(\ref{eq:ETH}) for many operators. The generalization to SU(N) gauge theories [a particular type of quantum field theories (QFTs) that is invariant under local gauge transformation] with their infinite number of degrees of freedom thus is possible if they are discretized, i.e., if one studies lattice gauge theories (LGTs). At the same time, the level spectrum must be dense for RMT and ETH to apply, implying the need for a large number of states and consequently significant computer power. Hence, we have to content ourselves to provide some numerical circumstantial evidence for selected thermalization properties of nonabelian gauge theories. Still, we believe that such limited evidence can be phenomenologically quite valuable. 

An example of thermalization in QCD (SU(3) gauge theory with dynamical quarks) is the production of a quark-gluon plasma as investigated at LHC and RHIC with great effort. For spin systems it has been demonstrated that the macroscopic equilibration time can be much smaller than the Thouless time controlling RMT behavior \cite{Wang:2021mtp}. This suggests that, e.g., a system formed by two colliding heavy ions is not characterized by a single thermalization time but possibly by a range of such times, depending on the observable that is being studied. Different equilibration time scales were indeed observed phenomenologically, leading researchers to postulate the existence of two distinct mechanisms, namely fast hydrodynamization and slow thermalization or phase space equilibration. As the relevant operators for hydrodynamics, i.e., local energy and momentum density, are of rather short range compared to the size of the thermalizing fireball, such a distinction is plausible, but an oversimplification in view of the discussion sketched above. An explicit demonstration of the domain size dependence of the thermalization time was presented in the context of a holographic (AdS/CFT) calculation \cite{Balasubramanian:2010ce, Balasubramanian:2011ur}, which showed for selected operators that the thermalization time scales linearly with the spatial support of the operator.

To study the properties of a QFT we follow \cite{Sharpe:2021wcq} and {\it define} a continuum SU(N) gauge theory as the limit of SU(N) LGT for vanishing lattice spacing and an infinite number of lattice points (the continuum and infinite volume limits). Current computational limitations force us to attempt to glean some properties of the full QFT from the properties of very small lattices. In spite of their small size, these lattices still involve Hilbert spaces with dimensions $10^5 - 10^6$.

The long success story of LGT has established beyond any reasonable doubt that the Wilson formulation of LGT gives correct results for a vast range of observables, reproduces the renormalization group running of the continuum theory in the ultraviolet (UV), and has a convergent infrared (IR) behavior for large volumes. The Hamiltonian formulation of Kogut and Susskind (KS) \cite{KS_Hamiltonian} is equally accepted as a valid formulation and forms the basis for the development of quantum computing for nonabelian gauge theories \cite{Banuls:2019bmf,Bauer:2023qgm}. Thus, it is sufficient to establish ETH behavior for these well-studied discretized formulations of SU(N) LGT, which can be done numerically, although the continuum limit is probably out of reach for currently available computing resources. In the past, this insight made it possible to establish RMT behavior for Euclidean gauge theories \cite{Berbenni-Bitsch:1997zmi,Halasz:1998qr, Verbaarschot:2000dy}. These studies demonstrated that RMT behavior could already be observed for quite small lattices, which makes us optimistic that a demonstration of ETH behavior for LGT is achievable.

This brings us to the question whether quantum or classical computers are optimal for such numerical studies. 
On one hand, the Wilson (Lagrangian) formulation can be implemented efficiently on classical computers using imaginary time propagation. Algorithms have been developed which allow for a limited description of real time evolution (see e.g., \cite{Ji:2013dva}). However, an attempt to establish the validity of Eq.~(\ref{eq:ETH}) within the Wilson framework would have to go far beyond the demonstrated realm of applicability of such methods. An alternative approach to study quantum chaotic behavior in matrix models and lattice gauge theory is the use of Gaussian density matrices with adjustable parameters \cite{Buividovich:2018scl,Gong:1993fz}.

On the other hand, very small systems could be insufficient to obtain the results we aim at and quantum computers or quantum simulators might be needed to fully profit from a Hamiltonian formulation. Our present calculations are performed on classical digital computers and correspondingly small systems. Our decision was primarily motivated by previous experience in establishing RMT behavior for gauge theories, which was also possible using very small lattices. Furthermore, classical ergodicity of SU(2) gauge theory could be demonstrated with clearly apparent finite-size scaling on rather small lattices of size $L^3$ for $L=2,4,6$ \cite{Bolte:1999th}. 

Presently we also limit ourselves to SU(2) gauge theory for which an efficient formulation in terms of angular momentum algebra was developed in \cite{Byrnes:2005qx,Zohar:2014qma,Klco:2019evd} and extended to SU(3) in \cite{Byrnes:2005qx,Ciavarella:2021nmj}. In addition, the studies reported here will be limited to lattices extending in one and two dimensions. The one-dimensional plaquette chain has been studied as a quantum system for very short chains \cite{Klco:2019evd,ARahman:2021ktn}. The honeycomb lattice was previously considered by \cite{Hayata:2023puo}, and point-splitting methods for square lattices were discussed in \cite{Zache:2023dko} as convenient approaches to implement the KS Hamiltonian in higher dimensions.  While the present study is still an exploratory one we hope that improved algorithms and larger computer resources will allow us to better control the continuum and infinite volume limits in the future.

For SU(2) there are three lines of investigation which we pursue in our present study within the constraints of the available computer resources: the dimensionality of the lattice, the number of plaquettes, and the  maximum angular momentum representation for electric gauge fields $\jmax$. To guarantee the exact reproduction of the KS Hamiltonian convergence in $\jmax$ must be demonstrated. Recently one of us (X.Y.) proposed a mapping of the 2+1 dimensional SU(2) gauge theory Hamiltonian onto an Ising system valid when the Hilbert space for each gauge link is constrained to the electric field representations $j=0,\frac{1}{2}$ \cite{Yao:2023pht}. This mapping allows one to find a complete set of energy eigenstates by direct diagonalization for chains of $N \lesssim 20$ plaquettes. While the simulated theory is thus not strictly a discretized version of SU(2) gauge theory, but rather a truncated model of it, the differences could be small, especially at strong coupling. This would be indicative of some universality of ETH behavior. 

Overall, as we will show, artifacts caused by the smallness of the studied systems are substantial, which limits the parameter range in which we can trust our results. Within these parameter ranges, however, our results clearly provide evidence for ETH behavior. However, we caution the reader at the outset that the question for which operators Eq.~(\ref{eq:ETH}) in a QFT should be valid is a highly non-trivial one. Physical operators typically renormalize multiplicatively such that we do not expect deviations from the generic ETH distribution properties. For unphysical operators, however, complications can occur such as operator mixing. The operators we study here are expected to have physical meaning in the continuum limit, which gives us hope that the ETH behavior observed on our small lattices survives in that limit.

We present three groups of results. First, in Section~\ref{sec:Ising_chain}, we focus on the system of a SU(2) plaquette chain with $j_{\rm max}=\frac{1}{2}$, which can be mapped onto an Ising chain. We study the statistical properties of matrix elements between different nearby energy eigenstates and show that their magnitudes form a Gaussian distribution. We also calculate the spectral function $f_{\cal A}(E,\omega)$ and study its dependence on energy $E$. Second, in Section~\ref{sec:Ising_honeycomb}, we use a similar mapping for a two-dimensional honeycomb lattice to calculate the complete set of eigenstates (within the truncated Hilbert space) for lattices of size $5\times 4$ (for periodic boundary conditions) and triangular lattices of 15 hexagonal plaquettes (for confining boundary conditions). For these we perform similar studies of their ETH properties. Third, in Section~\ref{sec:Convergence}, we explore the convergence of the energy spectrum with respect to the cutoff $\jmax$ in the electric field representation on small chains of three plaquettes. Also in this case, we search for the presence of ETH behavior in the energy range for which convergence of the basis is reached and investigate the corresponding spectral function $f_{\cal A}(E,\omega)$ in more detail. We finally summarize our results and discuss promising avenues of future research in Section~\ref{sect:conclusions}.

\section{Linear Plaquette Chain}
\label{sec:Ising_chain}
The discretized KS Hamiltonian of the 2+1 dimensional SU(2) gauge theory can be written as \cite{KS_Hamiltonian}
\begin{align}
    H=\frac{g^2}{2}\sum_{\rm links}(E^a_i)^2-\frac{2}{a^2g^2}\sum_{\rm plaquettes}\Box \,,
\label{eq:KS_Hamiltonian}
\end{align} 
where $g$ is the coupling constant, $a$ in the denominator the lattice spacing, $E^a_i$
the electric field operator along the direction $i=\hat{x}$ or $\hat{y}$ with the SU(2) index $a$ (both of which are implicitly summed) and $\Box \equiv {\rm Tr}[U^\dagger({\boldsymbol n},\hat{y}) U^\dagger({\boldsymbol n}+\hat{y},\hat{x}) U({\boldsymbol n}+\hat{x},\hat{y}) U({\boldsymbol n},\hat{x})]$ the plaquette operator at $\boldsymbol n = (n_xa, n_ya)$ that is the trace of the product of four link variables (Wilson lines) along a square lattice. (In Section~\ref{sec:Ising_honeycomb}, the honeycomb plaquette operator is defined as the trace of the product of six link variables and the prefactor of the magnetic term in the Hamiltonian will differ.) The coupling constant has the mass dimension of $[g]=0.5$; $E^2$ and $\Box$ are dimensionless operators. Throughout the article, we express all dimensionful quantities in units of the lattice spacing $a$ and imply the appropriate factor of powers of $a$ without denoting it explicitly. For example, when we state the coupling constant as $g^2=0.9$ and the energy as $E=2$, we imply $g^2=0.9a^{-1}$ ($g^2$ has dimension of energy in the 2+1 dimensional gauge theory) and $E=2a^{-1}$.

The discretized KS Hamiltonian can be represented in the electric energy basis, which labels the state on each link by the angular momentum quantum numbers $|jm_Lm_R\rangle$. They specify how the link state transforms under SU(2) gauge transformations generated by the electric fields at both the tail ($L$) and head ($R$). When the electric field and the link variable operators act on the corresponding link state, one has~\cite{Byrnes:2005qx,Zohar:2014qma}
\begin{align}
\label{eqn:Ebasis}
& \sum_{i,a}(E_i^a)^2 |jm_Lm_R\rangle = j(j+1)|jm_Lm_R\rangle \,, \nn \\
&\langle j'm_L'm_R' | U_{n_Ln_R} | j m_L m_R \rangle = \sqrt{({2j+1})/({2j'+1})} \nn\\
&  \qquad \times \langle j'\, m_L'| j \, m_L; \frac{1}{2}\, n_L \rangle \langle j\,  m_R; \frac{1}{2}\, n_R | j'\,  m_R' \rangle \,,
\end{align}
where $\langle j'\, m'| j \, m; J\, M \rangle $ denotes Clebsch-Gordan coefficients and $U_{n_Ln_R}$ denotes one entry of the SU(2) matrix. Physical states satisfy Gauss's law which in this lattice setup means all link states joining the same vertex transform as a SU(2) singlet. One can write out explicitly the matrix elements of the plaquette operator $\Box$ between physical states by using Eq.~\eqref{eqn:Ebasis} and Gauss's law, which has been done for a plaquette chain in \cite{Klco:2019evd,ARahman:2021ktn,ARahman:2022tkr} and a honeycomb lattice in \cite{Muller:2023nnk}. This allows one to explicitly write out the Hamiltonian matrix in the physical Hilbert space and exactly diagonalize it numerically.

We begin by further analyzing the statistical properties of the linear color-singlet plaquette chain with periodic boundary conditions on a truncated Hilbert space with $\jmax = \frac{1}{2}$ in the electric field representation. $j \leq \jmax$ denotes the SU(2) representation of the electric field operator on the lattice links. It was shown in \cite{Yao:2023pht} that the KS Hamiltonian for this system can be mapped onto the Hamiltonian for an Ising spin chain with nearest neighbor interaction
\be
H = \sum_{i=0}^{N-1} \left( J \sigma_i^z \sigma_{i+1}^z + h_z \sigma_i^z + h_x \sigma_i^x (i\sqrt{0.5})^{\sigma_{i-1}^z+\sigma_{i+1}^z}  \right) \, ,
\label{eq:H_Ising}
\ee
where $N$ denotes the number of plaquettes in the chain, and the coupling constants are related to the gauge coupling $g$ and lattice spacing $a$ as follows:
\be
J = -\frac{3g^2}{16}, \qquad h_z = -2J, \qquad h_x = \frac{1}{a^2g^2} \, .
\label{eq:couplings}
\ee
The periodic boundary condition is imposed by setting $\sigma_N^\alpha = \sigma_0^\alpha$. In the following we present results for $g^2 = 1.2$ and $N = 19$, except where stated otherwise. It is expected that in the large volume limit all nonzero values of $g^2$ are equivalent for the purpose of demonstrating the ETH, because the Hamiltonian is non-integrable for all nonvanishing $g^2$. However, for the small lattice sizes that are numerically accessible on classical computers, some choices of $g^2$ serve better to demonstrate the ETH behavior. Other choices will eventually exhibit the ETH behavior when the lattice becomes sufficiently large. Numerical evidence for this statement can be found in Appendix~\ref{app:g_dependence}, where we show the energy spectra in the case with $g^2=1$ for different chain length $N$.

The space of energy eigenstates for this system can be decomposed into 19 sectors with good linear momentum $pa = 2\pi k/N$ with integer $k \in \{ 0, 1, \cdots, N-1\}$. $k=i$ and $k=N-i$ correspond to two sectors that are related by the reflection. Each sector contains $(2^N-2)/N = 27,594$ states with the exception of the $k=0$ sector, which contains two additional states, the states with all the spins pointing downward or upward.

\begin{figure}[ht]
	\centering
 	\includegraphics[width=0.95\linewidth]{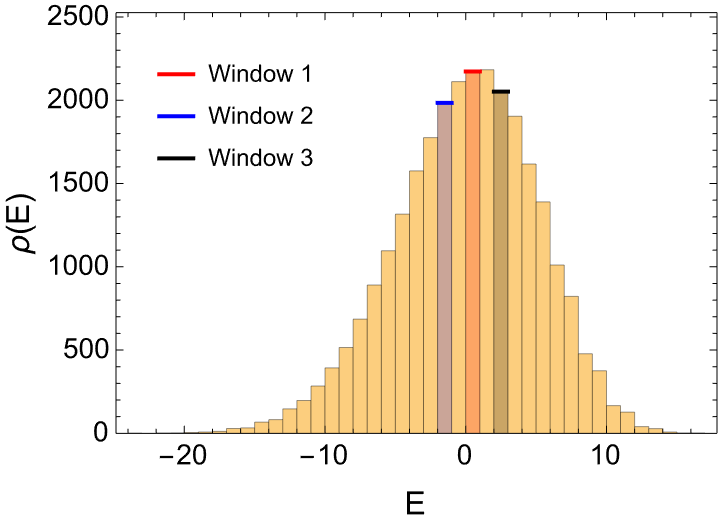}
	\caption{Histogram of the energy level density $\rho(E)$ (yellow bar) for the sector $k=1$ together with the three energy windows used in the statistical analysis of the off-diagonal matrix elements of the electric energy. Window 1: $0<E<1$ (center, red), Window 2: $-2<E<-1$ (left, blue), Window 3: $2<E<3$ (right, black).}
	\label{fig:EWindows}
\end{figure}

The energy level spacing statistics and the statistical properties of the matrix elements of the local 1-plaquette and 2-plaquette operators were studied in \cite{Yao:2023pht} with results that supported the assumption that the ETH is manifested in this system. Because these operators are not translation-invariant, they have nonvanishing matrix elements between states with different $k$. Here we consider matrix elements of the total electric energy operator given by the first two terms in the Hamiltonian Eq.~(\ref{eq:H_Ising})
\be
H_{\rm el} = \sum_{i=0}^{N-1} \left( J \sigma_i^z \sigma_{i+1}^z + h_z \sigma_i^z \right) \, ,
\label{eq:H_el}
\ee
which is translation-invariant and only couples states in the same momentum sector. In the following we focus on the $k=1$ sector; nearly identical results are obtained for other momentum sectors. (For the $k=0$ sector, the demonstration of Gaussian Orthogonal Ensemble (GOE) behavior in the level spacing statistics requires separation of the sector further into two subsectors with definite parity \cite{Yao:2023pht}. This will be done in Section~\ref{sec:Convergence} where we study the convergence of the spectrum under changes of the $\jmax$ cutoff for a short plaquette chain.) A histogram of the level density $\rho(E)$ is shown in Fig.~\ref{fig:EWindows}. The distribution has approximately the form $\rho(E) = A/\cosh^2[(E-\tilde{E})/\Delta]$ with $A=2250, \Delta=6.276, \tilde{E}=0.342$. The figure also shows three energy windows, which we will use for a statistical analysis of the off-diagonal matrix elements of the electric energy $H_{\rm el}$.

The distribution of level spacings was already studied in \cite{Yao:2023pht} and shown to be well described by the GOE distribution. As an additional sensitive statistical measure of the energy gap distribution, we have calculated the normalized energy gap ratio \cite{Oganesyan2007loc} 
\be
0 < r_\alpha = \frac{{\rm min}[\delta_\alpha,\delta_{\alpha-1}]}{{\rm max}[\delta_\alpha,\delta_{\alpha-1}]} \leq 1 \, ,
\label{eq:rgap}
\ee
where $\delta_\alpha = E_{\alpha+1} - E_\alpha$ is the energy gap between adjacent levels. The advantage of this measure is that it is not sensitive to the change of the level density with energy. The GOE prediction for this distribution is \cite{Jansen2019ETH}
\be
P_{\rm GOE}(r) = \frac{27}{4} \frac{r+r^2}{(1+r+r^2)^{5/2}}
\label{eq:rgap_GOE}
\ee
with the mean value $\langle r \rangle_{\rm GOE} \approx 0.5307$ for asymptotically large matrices \cite{Atas2013dist}.

\begin{figure}[ht]
	\centering
	\includegraphics[width=0.95\linewidth]{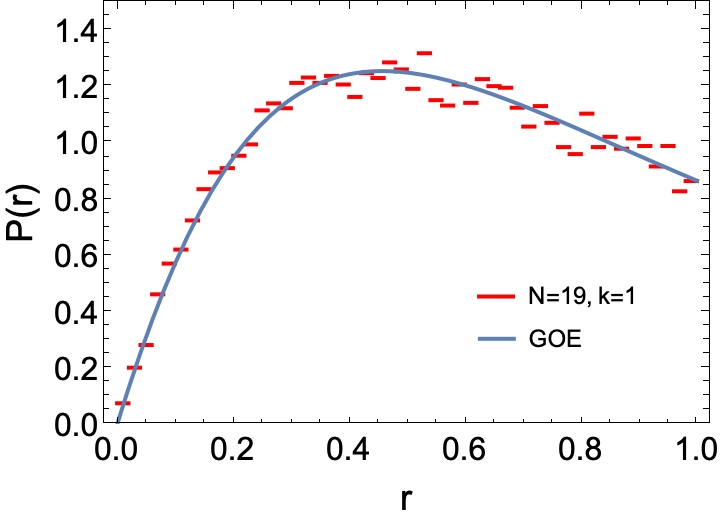}
	\caption{Histogram of the distribution of restricted energy gap ratios $r_\alpha$ (red horizontal bars), shown together with the GOE limit distribution $P_{\rm GOE}(r)$ (blue line).}
	\label{fig:rgap}
\end{figure}

In order to avoid distortions from the tails of the spectrum, we only include states in the range $(E_{\rm min}+E_{\rm av})/2 < E < (E_{\rm max}+E_{\rm av})/2$, where $E_{\rm min}$ and $E_{\rm max}$ are the lowest and highest energy eigenvalues, respectively, and $E_{\rm av}$ is the mean energy of the spectrum \cite{Jansen2019ETH}. Figure~\ref{fig:rgap} depicts our result for $P(r)$, shown as a histogram with red bars. The analytical distribution Eq.~(\ref{eq:rgap_GOE}), shown as a blue line, is seen to provide an excellent description, and the mean value $\langle r \rangle = 0.5340$ agrees well with the GOE prediction.

In order to investigate the statistics of off-diagonal matrix elements and calculate the spectral function $f_{\rm el}(E,\omega)$ for the electric energy operator, we select energy windows of width $\Delta E = 1$ shown in Fig.~\ref{fig:EWindows}. In the selection of the energy eigenstates for the study of matrix element statistics one has two options. One choice is to select energy pairs with a constraint on the mean, $|(E_\alpha + E_\beta)/2 - \bar{E}| < \Delta E/2$ where $\bar{E}$ is a chosen constant, the other is to require that both energies fall into a common window, $|E_\alpha - \bar{E}|, |E_{\beta} - \bar{E}| < \Delta E/2$. The first option corresponds to selecting a diagonal window in energy pair space, the second chooses a square window. In the limit $|\omega| = |E_\alpha-E_\beta| \ll \Delta E/2$ the two options effectively coincide. Here we choose the second option, because it is more closely related to the long-time behavior of a state defined by a wave packet with a narrow energy spread.

\begin{figure}[ht]
	\centering
    \includegraphics[width=0.95\linewidth]{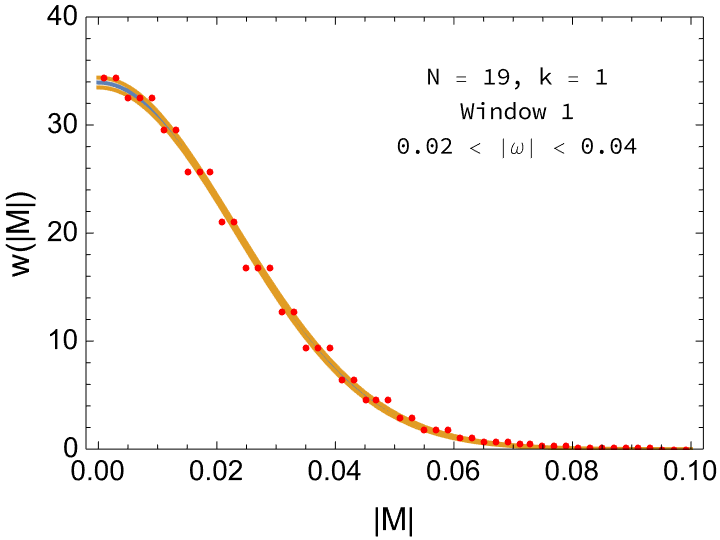}
	\caption{Distribution of matrix elements $|M_{\alpha\beta}|$ of the total electric energy operator $H_{\rm el}$ in the range $0.02 < |\omega| < 0.04$, shown together with a Gaussian fit.}
	\label{fig:ME_Gauss}
\end{figure}

We begin with the window $0 < E < 1$ (shown in red in Fig.~\ref{fig:EWindows} and corresponding to $\bar{E} = 0.5$, henceforth called Window 1), where the level density has its maximum. The window contains 2173 eigenstates. Figure~\ref{fig:ME_Gauss} shows a histogram of the values $|M_{\alpha\beta}| \equiv |\langle \alpha|H_{\rm el}|\beta\rangle|$ in the $\omega$ window $0.02 < |\omega| < 0.04$ together with a fit to a Gaussian distribution of the form
\be
w(|M|) = \sqrt{\frac{2}{\pi\sigma^2}} e^{-|M|^2/(2\sigma^2)} \, . 
\label{eq:ME_Gauss} 
\ee
Similarly good fits are obtained for other $\omega$ windows of the same width in the interval $|\omega| < 0.1$. 

\begin{figure}[ht]
	\centering
	\includegraphics[width=0.95\linewidth]{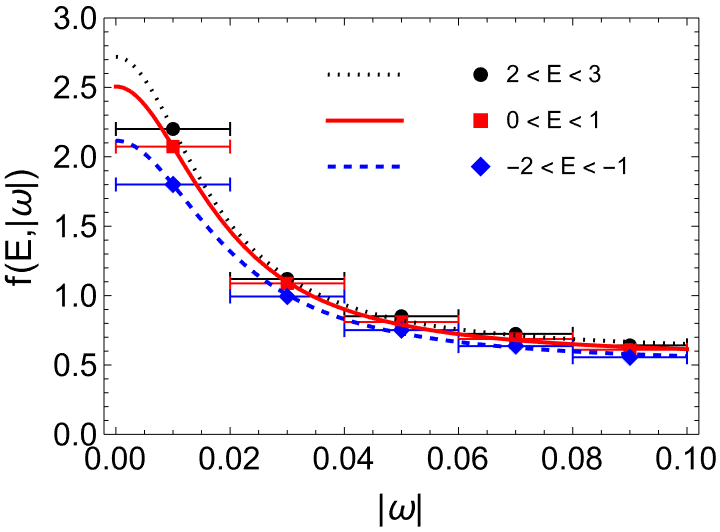}
	\caption{Spectral function $f_{\rm el}(E,\omega)$ together with their analytical fits of the form $a/(\omega^2+b^2) + c$ for three different energy windows as indicated in the legend: Window 1 (red, solid), Window 2 (blue, dashed), Window 3 (black, dotted).}
	\label{fig:spect_fct}
\end{figure}

The width of the Gaussian is related to the spectral function $f_{\rm el}(E,\omega)$. Using the ETH relation Eq.~(\ref{eq:ETH}), ${\rm Tr}[{\boldsymbol R}^2] \equiv R_{\alpha\beta}R_{\beta\alpha} = 1$, and $e^{S(E)} = \rho(E)$ one finds
\be
\sigma^2 = {\rm Tr}[{\boldsymbol M}^2] = \frac{f_{\rm el}(E,\omega)^2}{\rho(E)} \, , 
\label{eq:M^2}
\ee
which implies $f_{\rm el}(E,\omega) = \sigma(\omega)\sqrt{\rho(E)}$. Figure~\ref{fig:spect_fct} shows the spectral function $f_{\rm el}(E,\omega)$ for $0 < |\omega| < 0.1$ in three energy windows: $0 < E < 1$ (Window 1, red), $-2 < E < -1$ (Window 2, blue), and $2 < E < 3$ (Window 3, black). For small values of $|\omega|$ the spectral function is well described by the functional form 
\be
f(E,\omega) = \frac{a}{\omega^2+b^2} + c
\label{eq:peak_fit}
\ee 
representing a diffusive transport peak superimposed on a flat pedestal. The fits are shown as solid, dashed, and dotted lines, respectively. We shall discuss the shape of the spectral function in more detail and over a wider $\omega$ range in Section~\ref{sec:Convergence}.

The final statistical test of off-diagonal matrix elements we perform is to check whether they form a GOE. It is also worth noting that, although the matrix elements are complex numbers for energy eigenstates with definite momentum, the associated Gaussian ensemble for comparison is the GOE (and not the Gaussian Unitary Ensemble, GUE), because the Hamiltonian is time reversal invariant. We calculate the second and fourth moments of the band matrix obtained by keeping only matrix elements between eigenstates that differ in energy by less than $\Delta E(T) = 2\pi/T$ \cite{Wang:2021mtp}. The idea is that contributions to matrix elements at time $T$ between wave packets from eigenstate pairs $\alpha,\beta$ with $|E_\alpha-E_\beta| > \Delta E(T)$ cancel out because of their random energy phases, but the randomness of matrix elements between states differing by less than $\Delta E(T)$ may be required for some thermal properties to manifest themselves. As time progresses, only the contributions from an increasingly narrower band of energy eigenstate pairs are relevant. 

We define the projected submatrix elements of an operator ${\cal O}$ as
\be
{\cal O}^T_{\alpha\beta} = \left\{
\begin{array}{cl}
\langle \alpha|{\cal O}| \beta\rangle, & |E_\alpha-E_\beta | \leq \Delta E(T) \\
0, & |E_\alpha-E_\beta | > \Delta E(T) \,,
\end{array} \right. 
\label{eq:O^T}
\ee
where the diagonal part is always included. Our definition differs slightly from the one adopted in \cite{Wang:2021mtp} in that we consider all pairs of states within each of the fixed ($T$-independent) energy windows shown in Fig.~\ref{fig:EWindows}. As $T$ increases, the matrix ${\cal O}^T$ contains a growing number of vanishing elements. The GOE measure $\Lambda^T$ is defined as \cite{Wang:2021mtp}
\be
\label{eqn:lambdaT}
\Lambda^T =\frac{\left(\Tr[({\cal O}_c^T)^2]\right)^2}{d\,\left(\Tr[({\cal O}_c^T)^4]\right)} \, ,
\ee
where $d$ is the number of eigenstates in the chosen energy window and ${\cal O}_c^T = {\cal O}^T - \Tr[{\cal O}^T]/d$ is the traceless part of the matrix. In our case ${\cal O}_c^T$ is complex but Hermitian, which ensures that $\Tr[({\cal O}_c^T)^n]$ is real. For matrices approaching a perfect GOE at late times, one expects $\Lambda^T \to \frac{1}{2}$. 

The $\Lambda^T$ measure stands out by being a sensitive test not only of the Gaussian distribution of the absolute values of the matrix elements, but also for correlations among their signs, which encode the orthogonality of the GOE matrix ensemble. In practice, as a smaller and smaller number of elements of the projected matrix ${\cal O}^T$ is nonvanishing for $T \to\infty$, the statistical significance of the result for $\Lambda^T$ deteriorates. This can be seen if one arbitrarily replaces many of the elements of a large matrix from the GOE by zero, converting it into a band matrix with sparse off-diagonal elements.

\begin{figure}[ht]
	\centering
	\includegraphics[width=0.95\linewidth]{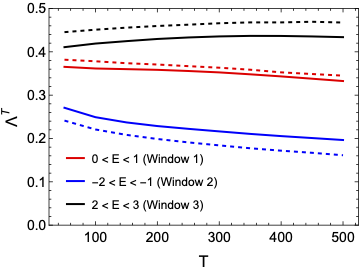}
	\caption{The GOE measure $\Lambda^T$ for the projected matrix elements Eq.~(\ref{eq:O^T}) of the electric energy operator as function of time $T$ for eigenstates in Window 2 (bottom), Window 1 (middle) and Window 3 (top). Solid lines: $N=19$; dashed lines: $N=17$.}
\label{fig:LambdaT19}
\end{figure}

Our results for the three energy windows for $N=17$ (dashed lines) and $N=19$ (solid lines) are shown in Fig.~\ref{fig:LambdaT19}. As can be seen, the value of $\Lambda^T$ increases with the energy of the window. The values $\Lambda^T \approx 0.45 - 0.47$ closest to the GOE limit 0.5 are reached for Window 3, for which the deviation from the GOE value 0.5 is compatible with those expected for a sparse off-diagonal GOE matrix. However, the deviations for the the other two energy windows are too large to be explained in this way. This means that the off-diagonal matrix elements of the operator $H_{\rm el}$ in these energy windows do not form GOEs although the distribution of the absolute values is Gaussian. We attribute this to the fact that the electric energy operator in the $\jmax = \frac{1}{2}$ truncated basis is constrained to have values that are multiples of $j_{\rm max}(j_{\rm max}+1)=\frac{3}{4}$. We shall see in Section \ref{sec:Convergence} that the off-diagonal matrix elements of $H_{\rm el}$ are closer to a GOE when $j_{\rm max}$ is increased, and we expect this property to be realized in the continuum limit where the spectrum of the electric energy operator becomes continuous. It will be interesting to explore whether the matrix elements of other operators exhibit a faster approach to the GOE.

\section {Honeycomb Lattice}
\label{sec:Ising_honeycomb}

In this section we will generalize the previous one-dimensional (1D) plaquette chain study to two dimensions (2D). The new dissipative mode that contributes to thermalization in 2D, which does not exist in 1D, is the shear mode. Scale invariance, which is exact in gauge theories at the classical level and only broken at the quantum level by the trace anomaly, constrains dissipation of certain modes. This manifests itself by the strong suppression of bulk viscosity in gauge theories at high temperature. On the other hand, dissipation in the shear channel is not suppressed by symmetry and remains finite at strong coupling.
Therefore, the new physical insight one can gain from the ETH study in 2D is the thermalization dynamics related to shear viscosity. 

\begin{figure}[ht]
\subfloat[Parallelogram.\label{fig:honeycomb_rectangle}]{%
  \includegraphics[height=1.4in]{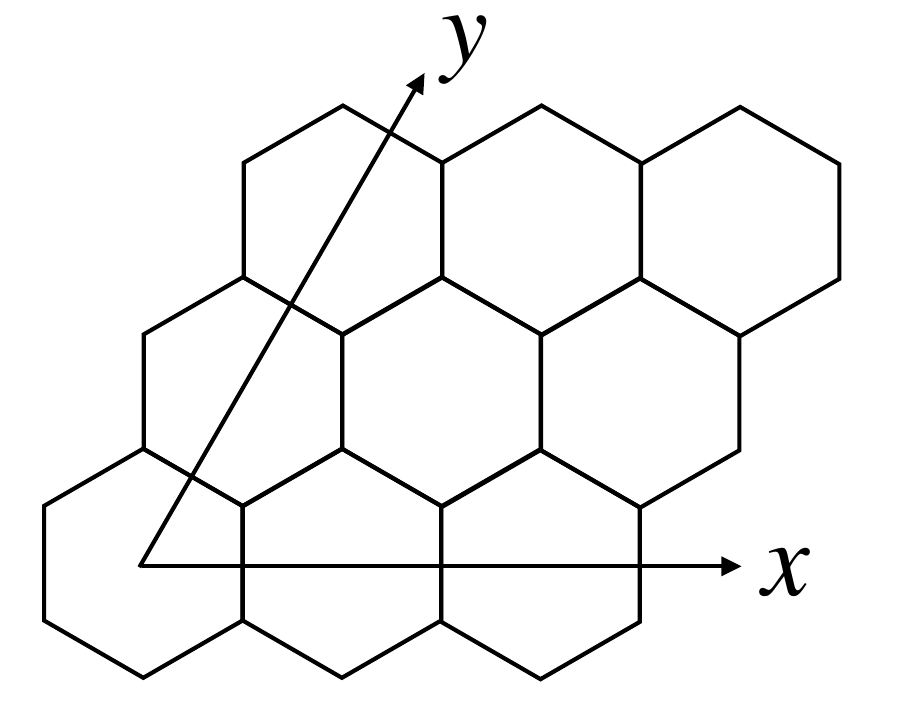}%
}\hfill
\subfloat[Triangle.\label{fig:honeycomb_triangle}]{%
  \includegraphics[height=1.4in]{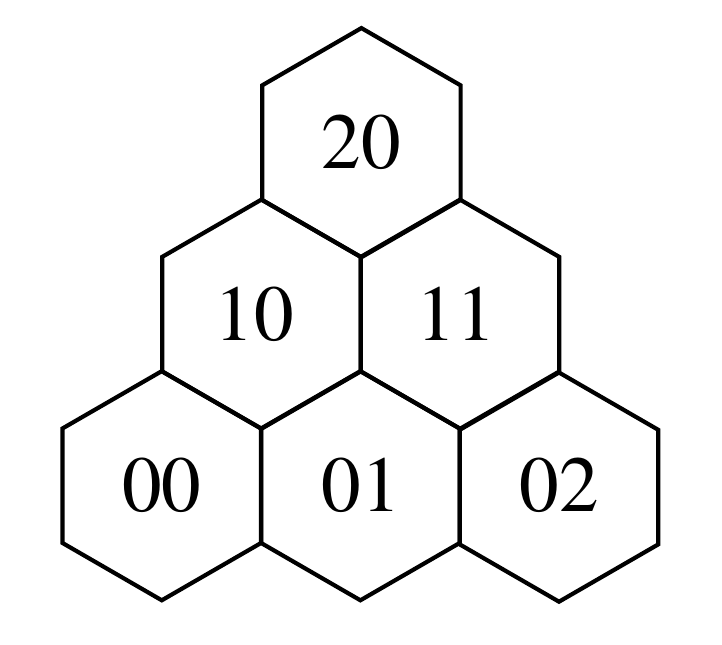}%
}
\caption{Examples of the honeycomb lattice considered in this work: (a) a parallelogrammatic shape with $N_x=N_y=3$; (b) a triangular shape with $N=3$.}
\label{fig:honeycomb_lattice}
\end{figure}

We will present the results obtained for the honeycomb lattice with $\jmax=\frac{1}{2}$. At each vertex, physical states can be fully specified by the $j$ values on the links connecting to the vertex if the vertex has fewer than four links. If the vertex has four links connected or more, as on a square lattice in 2D (spatial dimensions) or a cubic lattice in 3D, physical states depend on the order of how the color representations on different links are coupled and thus cannot be fully specified by these $j$ values alone. This motivates the use of the honeycomb lattice where each vertex is touched by at most three links. Two examples of honeycomb lattice setups are shown in Fig.~\ref{fig:honeycomb_lattice}.

The Hamiltonian of the 2D SU(2) gauge theory on a honeycomb lattice with $\jmax=\frac{1}{2}$ has been mapped onto a 2D spin model in \cite{Muller:2023nnk} for both periodic and closed boundary conditions. We will first consider the periodic boundary condition before discussing the closed boundary condition.

\subsection{Periodic Boundary Condition}
\label{subsec:Periodic Boundary Condition}
The spin Hamiltonian for the original SU(2) lattice gauge theory with $\jmax=\frac{1}{2}$ on a parallelogram under the periodic boundary condition can be written as~\cite{Muller:2023nnk}
\begin{align}
\label{eq:H_periodic}
H &= \sum_{(i,j)} \Big( J \sigma^z_{i,j}(\sigma^z_{i+1,j} + \sigma^z_{i,j+1} + \sigma^z_{i+1,j-1}) \nn\\
& \qquad\quad + h_x (-0.5)^{c_{i,j}} \sigma^x_{i,j} \Big)\,,
\end{align}
where $J=-\frac{9\sqrt{3}g^2}{32}$, $h_x=\frac{4\sqrt{3}}{9a^2g^2}$ and
\begin{align}
c_{i,j} & = \Pi^+_{i,j+1}\Pi^-_{i+1,j} + \Pi^+_{i+1,j}\Pi^-_{i+1,j-1} + \Pi^+_{i+1,j-1}\Pi^-_{i,j-1} \nn\\
& + \Pi^+_{i,j-1}\Pi^-_{i-1,j} 
+ \Pi^+_{i-1,j}\Pi^-_{i-1,j+1} + \Pi^+_{i-1,j+1}\Pi^-_{i,j+1} \nn\\[5pt]
\Pi^\pm_{i,j} &= \frac{1 \pm \sigma^z_{i,j}}{2} \,.
\end{align}
The $(i,j)$ symbol labels the plaquette position at $j$ along the $x$-direction defined by $(1,0)$ in the Cartesian 2D plane and $i$ along the $y$-direction defined by $(\frac{1}{2}, \frac{\sqrt{3}}{2})$, as shown in Fig.~\ref{fig:honeycomb_rectangle}. We assume there are $N_x$ plaquettes along the $x$-direction and $N_y$ along the $y$-direction.
The first term of Eq.~\eqref{eq:H_periodic} is the electric part of the Hamiltonian while the second term is the magnetic part. The magnetic part involves an exponential of Pauli matrices but one can rewrite it purely in terms of tensor products of Pauli matrices~\cite{Muller:2023nnk}. In our current studies using classical computers, both work equally well. For a quantum simulation, the latter format is more convenient to use since it can be easily mapped onto qubits.

Under the periodic boundary condition, the Hamiltonian in Eq.~\eqref{eq:H_periodic} is translationally invariant along the $x$ and $y$ directions by one lattice unit defined above: $[H,\hat{T}_x] = [H,\hat{T}_y] = 0$. As a result, the Hamiltonian and the translational operators can be simultaneously diagonalized. Each eigenstate belongs to a particular momentum sector given by $2\pi k_x/N_x, 2\pi k_y/N_y$ where $k_x\in[0,1,\cdots, N_x-1]$ and $k_y\in[0,1,\cdots, N_y-1]$. The Hamiltonian in each momentum sector can be constructed as discussed in \cite{Muller:2023nnk} and diagonalized exactly for small lattice sizes.

\begin{figure}
\subfloat[Eigenenergy density.\label{fig:eigenspectrum_2d_a}]{%
  \includegraphics[height=1.25in]{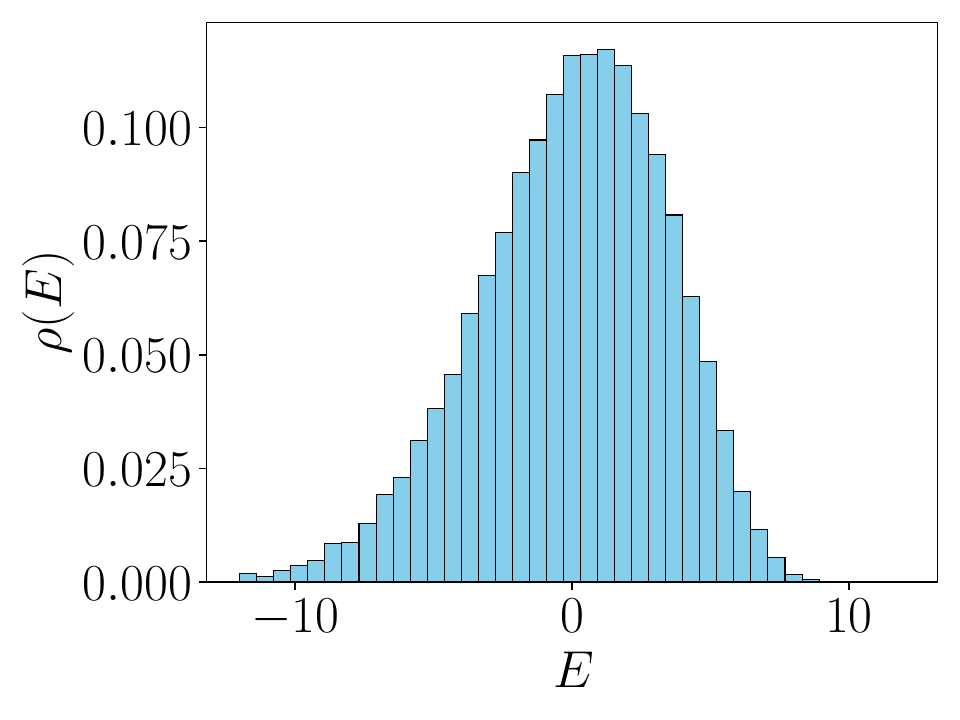}%
}\hfill
\subfloat[Gap.\label{fig:eigenspectrum_2d_b}]{%
  \includegraphics[height=1.25in]{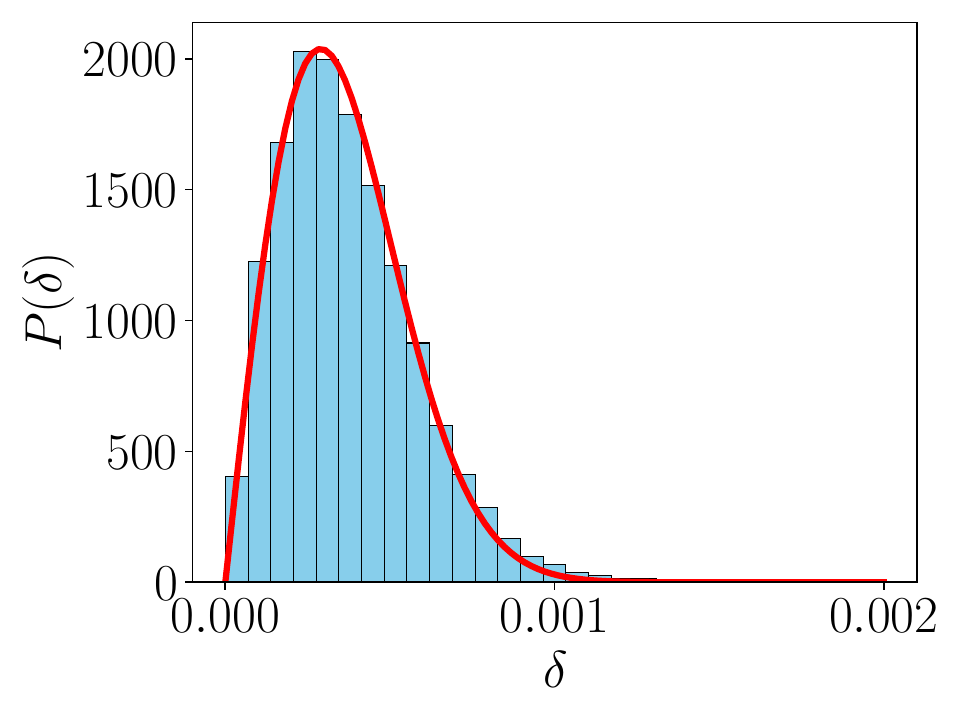}%
}

\subfloat[Rescaled gap.\label{fig:eigenspectrum_2d_c}]{%
  \includegraphics[height=1.25in]{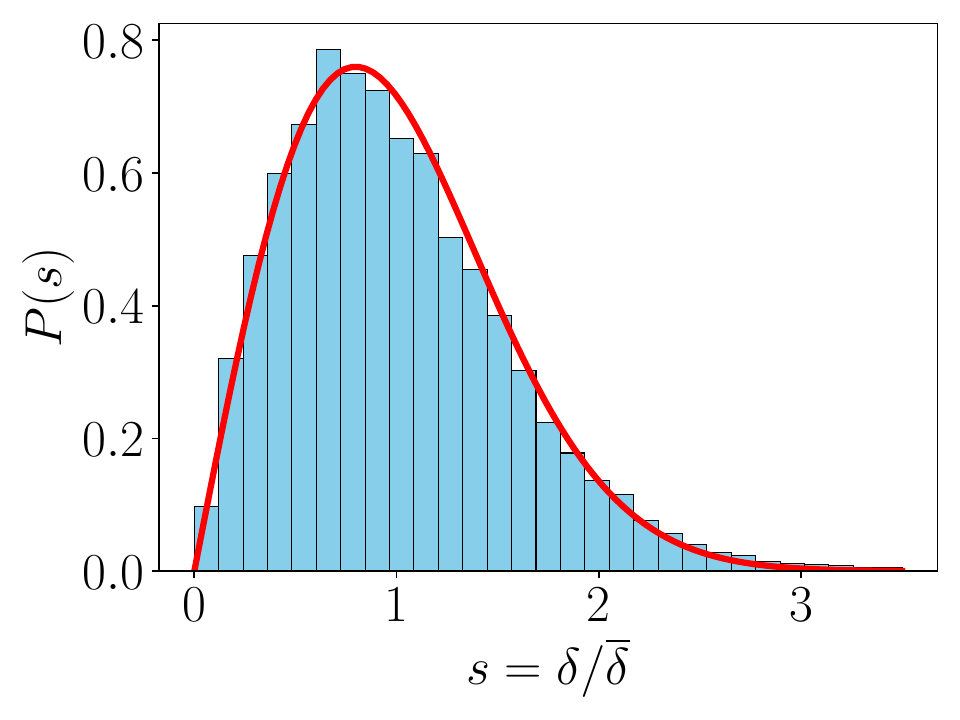}%
}\hfill
\subfloat[Restricted gap ratio.\label{fig:eigenspectrum_2d_d}]{%
  \includegraphics[height=1.25in]{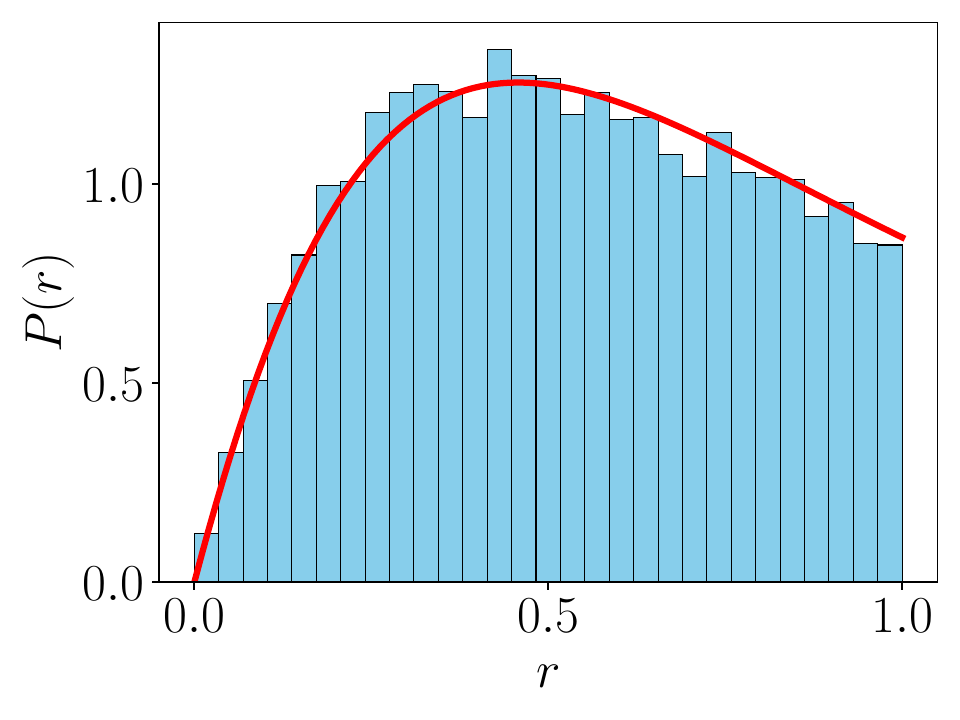}%
}
\caption{Density of eigenstates, distributions of gaps, rescaled gaps and gap ratios in the momentum $k_x=k_y=1$ sector on the $N_x=5,N_y=4$ lattice for $g^2=0.75$. The red curve in (b) is a Wigner surmise fit and the other two red curves in (c) and (d) are GOE predictions.}
\label{fig:eigenspectrum_2d}
\end{figure}

We first study the statistics of the eigenspectrum. We consider the $k_x=1,k_y=1$ sector in the $N_x=5, N_y=4$ system, which contains 26163 states. (We choose $N_x \neq N_y$ to avoid discrete hexagonal symmetry.) We plot the density of eigenstates in Fig.~\ref{fig:eigenspectrum_2d} for $g^2=0.75$, where we also plot the distributions of the energy gaps $\delta$ between nearest eigenstates and the rescaled gaps $s\equiv \delta/\overline{\delta}$ ($\overline{\delta}$ is the average of the gap), and the distribution of the restricted gap ratios defined in Eq.~\eqref{eq:rgap}, for the eigenstates in the middle of the energy spectrum (we removed the first 5000 and last 5000 eigenstates when generating these three plots). The red curve in the gap distribution Fig.~\ref{fig:eigenspectrum_2d_b} is a fit from the Wigner surmise $P_{\rm ws}(\delta) = a\delta \exp(-b \delta^2)$. The fitted parameter values are $a\approx 10164541, b\approx 5337217$. Level repulsion is clearly seen in the gap distribution, which is a feature of a quantum system whose classical counterpart is chaotic. 

We emphasize that level repulsion is very sensitive to any discrete symmetry of the system. If we plot the gap distribution for momentum sectors involving a zero momentum (i.e., $k_x=0$ and/or $k_y=0$), level repulsion cannot be seen and the distribution does not exhibit a Wigner-Dyson shape. This is due to the remaining parity symmetry in the zero momentum sectors. If we further separate each zero momentum sector into a parity-even and parity-odd sector and plot the gap distribution in each parity sector, we will see level repulsion clearly again and the distribution can be well fitted by a Wigner-Dyson shape, as shown for the chain case in \cite{Yao:2023pht}. In Fig.~\ref{fig:eigenspectrum_2d_c}, the red curve is a prediction from the GOE of $2\times2$ random matrices which states
\begin{align}
P(s) = \frac{\pi s}{2}\exp \Big(-\frac{\pi s^2}{4} \Big) \,.\label{eq:Wigner_Dyson_distribution}
\end{align}
In Fig.~\ref{fig:eigenspectrum_2d_d}, the red curve is a GOE prediction as written in Eq.~\eqref{eq:rgap_GOE}. The expectation value of the restricted gap ratio can be calculated as $\langle r \rangle\approx0.5284$, which is very close to the GOE prediction $\langle r \rangle_{\rm GOE}\approx0.5307$.
As can be seen, the statistics of eigenenergies in the middle of the spectrum can be well described by the GOE, up to statistical uncertainties due to the finite size.

Before moving on to operator matrix elements, we want to comment on the effect of $g^2$ values on the energy spectrum. Here we focus on the case with $\jmax = \frac{1}{2}$. In Section~\ref{sec:Convergence}, we will discuss how to choose $\jmax$ with varying values of $g^2$ in order to obtain physical results. In the large volume limit, it is expected that any value of $g^2$ will lead to qualitatively similar level statistics that are well described by the GOE. However, with a finite lattice size that is numerically accessible by current classical computers, some choices of $g^2$ values will give better results than others. For example, the reasonably good results shown above are obtained from the choice $g^2=0.75$. If we choose $g^2=1$, we observe a spiky structure in the level density, which distorts the distributions of gaps and gap ratios away from the GOE predictions. We believe these are finite volume effects and expect them to go away as the lattice size increases. Numerical evidence for this expectation is shown in Appendix~\ref{app:g_dependence} for the plaquette chain (we choose the chain to demonstrate this since we can go to a relatively large lattice size in the quasi-1D case such that momentum level separation is smaller and so is relevant kinetic energy).

\begin{figure}
\subfloat[\label{fig:He_off_2d_a}]{
\includegraphics[height=1.2in]{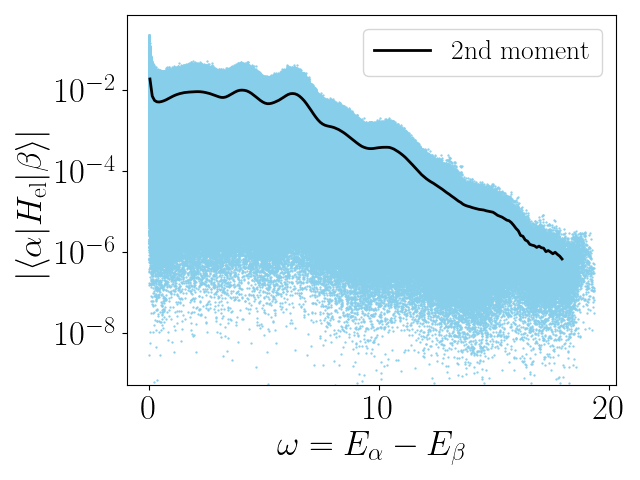}}
\hfill
\subfloat[\label{fig:He_off_2d_b}]{
\includegraphics[height=1.2in]{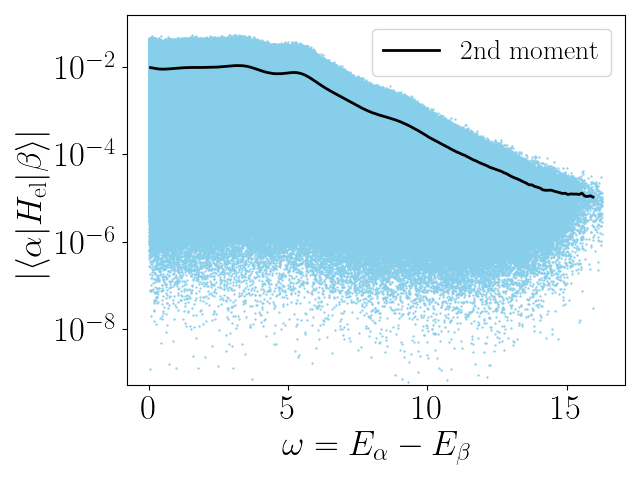}}
\\
\subfloat[\label{fig:He_off_2d_c}]{  \includegraphics[height=1.2in]{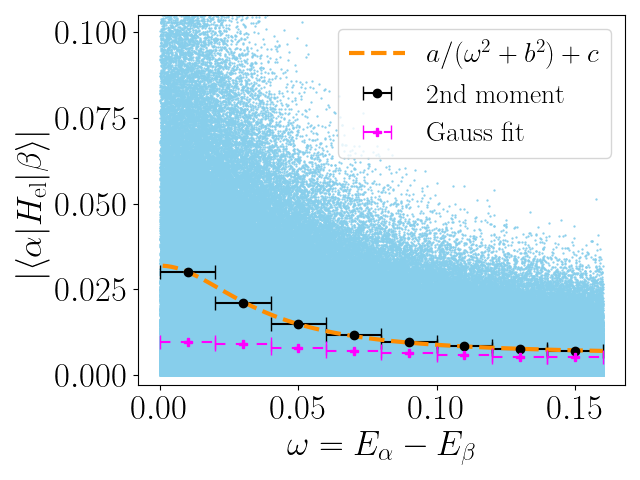}}
\hfill
\subfloat[\label{fig:He_off_2d_d}]{
\includegraphics[height=1.2in]{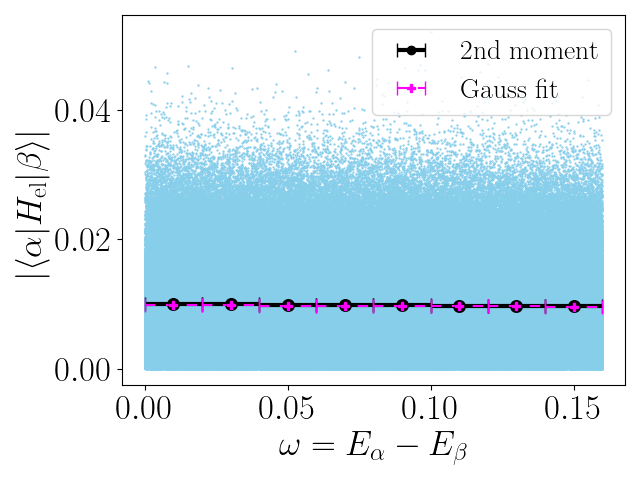}}
\caption{Off-diagonal matrix elements of the electric energy in the eigenbasis in the $k_x=k_y=1$ momentum sector on the $N_x=5,N_y=4$ lattice for $g^2=1$ (left) and $g^2=0.75$ (right). The plots in the top row are in the full $\omega$ region while the bottom row is an enlargement of the small $\omega$ region. The black curves are the widths calculated by the second moment of the matrix element distribution. The magenta curves are the widths fitted by Gaussian distributions. The yellow curve is a fit of the second moment results in the $g^2=1$ case with $a\approx3.50\times10^{-5}$, $b^2\approx1.34\times10^{-3}$ and $c\approx5.77\times10^{-3}$.}
\label{fig:He_off_2d}
\end{figure}

Next we study the matrix elements of the electric energy in the energy eigenbasis. The electric energy is given by
\begin{align}
H_{\rm el} = \sum_{(i,j)}  J \sigma^z_{i,j}(\sigma^z_{i+1,j} + \sigma^z_{i,j+1} + \sigma^z_{i+1,j-1}) \,,
\end{align}
which is translationally invariant along the $x$ and $y$ directions. As a result, the off-diagonal matrix elements of $H_{\rm el}$ vanish between states in different momentum sectors. In the following, we will only compute the matrix elements of $H_{\rm el}$ in one momentum sector.

We consider the $k_x=k_y=1$ momentum sector on the $N_x=5,N_y=4$ lattice and plot the magnitudes of the off-diagonal matrix elements, i.e., $|\langle \alpha| H_{\rm el}|\beta \rangle|$ in Fig.~\ref{fig:He_off_2d} for states satisfying $2<E_\alpha+E_\beta<4$ for two choices of couplings: $g^2=1$ and $g^2=0.75$. Since the distributions are symmetric around $\omega\equiv E_\alpha-E_\beta=0$, we only plot the distributions for $\omega>0$. If the ETH applies, the distribution of the off-diagonal matrix elements in a small $\omega$ window should be Gaussian when the number of elements is large. We test this in the small $\omega$ region by comparing the width obtained in two ways. In the first way, we fit the random variable distribution in an $\omega$ window of size $0.02$ to a Gaussian to obtain the width, shown in magenta. Alternatively, we calculate the second moment
\begin{align}
\frac{1}{N_{\rm terms}} \sum_{\alpha,\beta} |\langle \alpha| H_{\rm el} | \beta \rangle |^2  \,\label{eq:second moment}
\end{align}
of the matrix constrained to pairs of states with $\omega_{\rm min} < \omega <\omega_{\rm max}$, shown in black ($\omega_{\rm min}$ and $\omega_{\rm max}$ are the lower and upper bounds of an $\omega$ window). If the off-diagonal matrix elements magnitudes satisfy a Gaussian distribution, the two methods will give the same result. In the case of $g^2=1$ (Fig.~\ref{fig:He_off_2d_c}) we see a noticeable disagreement between the two methods for small $\omega$, which indicates the magnitudes of the off-diagonal matrix elements do not exactly follow Gaussian distributions. On the other hand, we see a very good agreement between the methods in the case of $g^2=0.75$ (Fig.~\ref{fig:He_off_2d_d}), which shows the magnitudes of the off-diagonal matrix elements in the small $\omega$ region follow Gaussian distributions. The contrast between the two cases of different couplings is consistent with the above studies of the level statistics, where we see the case with $g^2=0.75$ can be well described by the GOE.

We also look at the off-diagonal matrix elements in a wider $\omega$ region. As can be seen, the distributions shown in Figs.~\ref{fig:He_off_2d_a} and~\ref{fig:He_off_2d_b} exhibit bumpy plateaus at small and intermediate $\omega$ and decay exponentially at large $\omega$. The case with $g^2=1$ shown in Fig.~\ref{fig:He_off_2d_a} contains more bumpy structures, in particular at small $\omega$ as shown in Fig.~\ref{fig:He_off_2d_c}, where we fit the $\omega$ dependence of the second moment extracted width by the function $a/(\omega^2+b^2) + c$. The peak near $\omega\approx0$ is likely related to diffusive transport processes. However, as we decrease the coupling to $g^2=0.75$, we find the peak disappears, as shown in Fig.~\ref{fig:He_off_2d_d}. The reason for the disappearance of the transport peak at weaker coupling is unclear and may be a small lattice artifact.

\begin{figure}[t]
\centering
\includegraphics[height=2.3in]{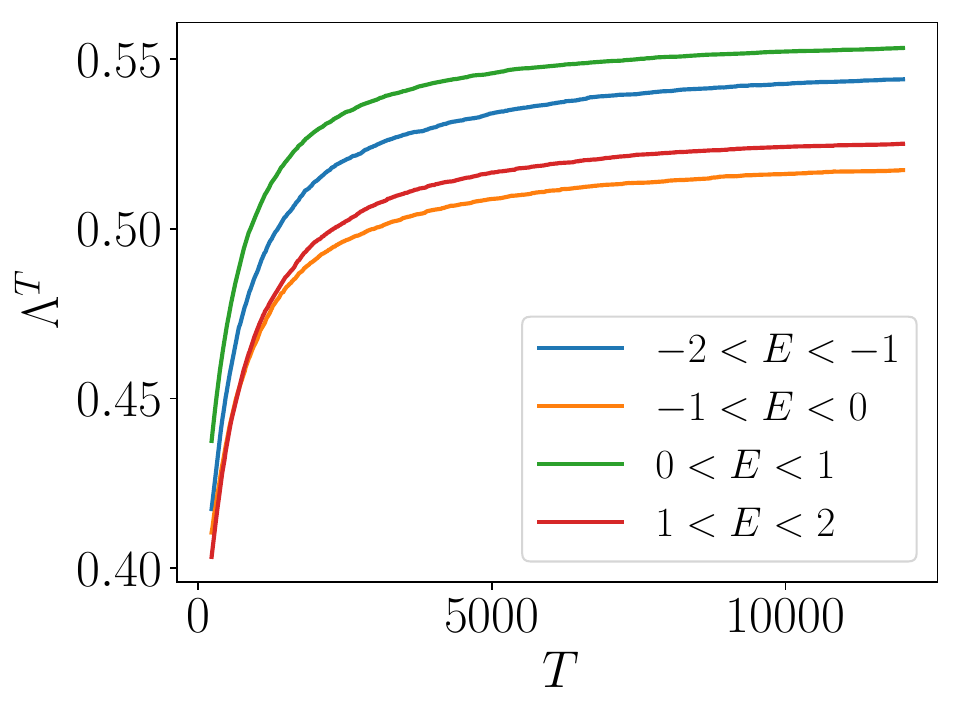}
\caption{$\Lambda^T$ observable as a function of $T$ for the electric energy operator in four eigenenergy windows $-2<E<-1$, $-1<E<0$, $0<E<1$ and $1<E<2$ in the $k_x=k_y=1$ sector on the $N_x=5,N_y=4$ lattice with $g^2=0.75$. (The matrix still has 7944 nonzero elements at $T=10000$.)}
\label{fig:lambdaT_2d}
\end{figure}

We next study the GOE measure $\Lambda^T$ defined in Eq.~\eqref{eqn:lambdaT}, which tracks the approach of the off-diagonal part of the matrix to GOE behavior. We choose four energy windows in the case of $g^2=0.75$ for our analysis: $-2<E<-1$, $-1<E<0$, $0<E<1$ and $1<E<2$. The results are shown in Fig.~\ref{fig:lambdaT_2d}. If the off-diagonal matrix elements are described by the GOE, the value of $\Lambda^T$ will reach 0.5. We see some deviations from the GOE prediction, which indicates the off-diagonal matrix elements are still correlated in their signs. (We have shown in Fig.~\ref{fig:He_off_2d} that the magnitudes of the off-diagonal matrix elements satisfy Gaussian distributions in the small $\omega$ region. So the deviations seen here reflect the sign correlations and thus non-orthogonality rather than the non-Gaussianity.) It is worth further investigating the $\Lambda^T$ observable on bigger lattices in future work.

\begin{figure}[t]
\centering
\includegraphics[width=0.95\linewidth]{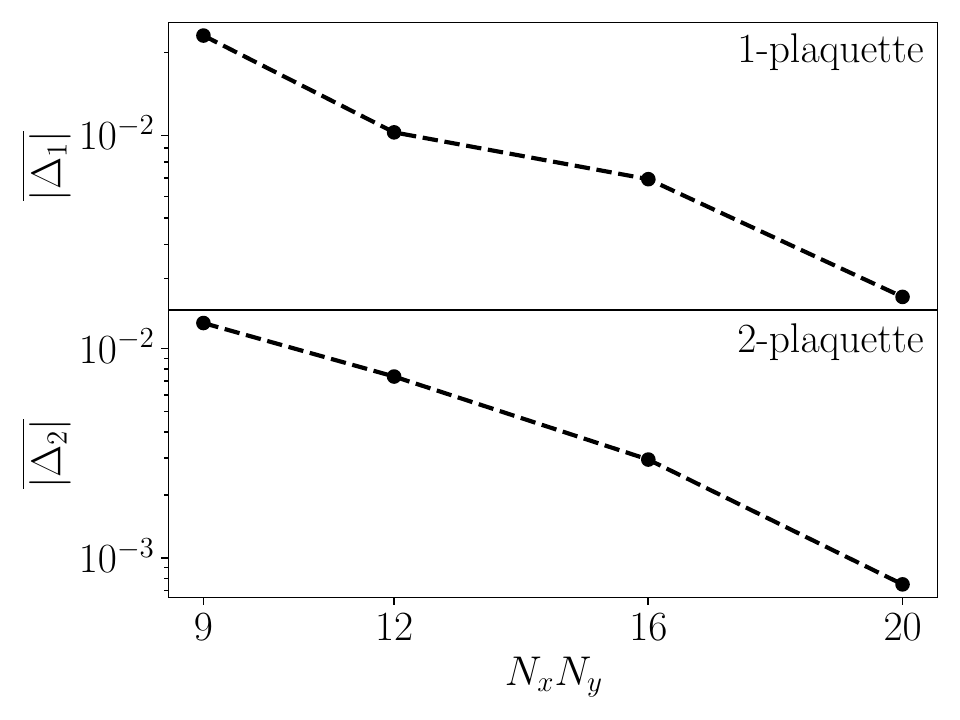}
\caption{Averaged magnitude of the difference between the diagonal matrix element and the microcanonical ensemble average decays roughly exponentially as a function of system size for both Wilson loop operators.}
\label{fig:diag_2d_pbc}
\end{figure}

Finally, we study matrix elements of operators that break the translational invariance. Following \cite{Yao:2023pht}, we study Wilson loops corresponding to 1-plaquette $O_1$ and 2-plaquette $O_2$ operators. The Pauli matrix representation of these Wilson loop operators and their matrix elements in the momentum basis have been worked out for the honeycomb lattice in \cite{Muller:2023nnk}. We use all the momentum sectors up to (including) $k_x=\lfloor N_x/2 \rfloor$ and $k_y=\lfloor N_y/2 \rfloor$ in the $g^2=1$ case. We first investigate lattice size dependence of the diagonal part of their matrix elements by considering lattices of size $3\times 3, 4\times 3, 4\times 4$, and $5\times 4$. We estimate the deviation of the diagonal part from the microcanonical ensemble average by using a proxy for the microcanonical ensemble made up of 10 eigenstates below and 10 above the eigenstate under consideration~\cite{Yao:2023pht}. The deviation is estimated as
\begin{align}
\Delta_i(\alpha) \equiv \langle \alpha| O_i |\alpha\rangle - \frac{1}{21}\sum_{\beta=\alpha-10}^{\alpha+10} \langle \beta| O_i |\beta\rangle \,.
\end{align}
We plot the magnitude of $\Delta_i$ averaged over all eigenstates except for the 20 lowest and 20 highest states in Fig.~\ref{fig:diag_2d_pbc}, as a function of lattice size. We find that $\overline{|\Delta_i|}$ decreases  approximately exponentially with lattice size, confirming the exponentially decaying factor $e^{-S/2}$ in the fluctuating part of Eq.~\eqref{eq:ETH} for most eigenstates, since the entropy is an extensive quantity, $S\propto N_xN_y$. In other words, we demonstrated the diagonal matrix elements of the two Wilson loop operators are exponentially close to the microcanonical ensemble average value, as the system size increases.

\begin{figure}[t]
\centering
\includegraphics[height=2.3in]{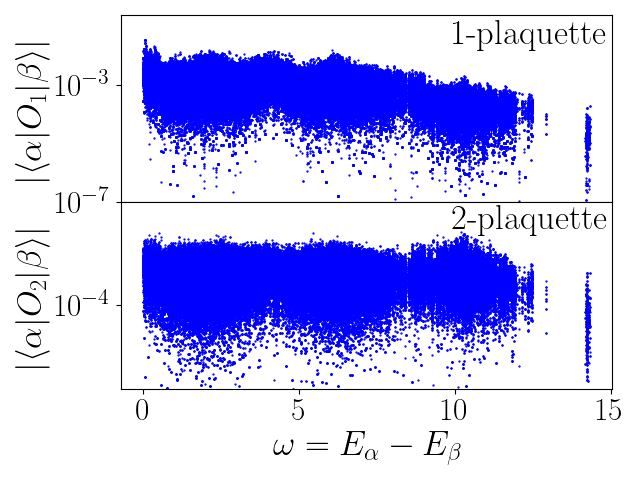}
\caption{Off-diagonal matrix elements of Wilson loop operators in a small energy shell.}
\label{fig:off_2d_pbc}
\end{figure}

We also plot the off-diagonal matrix elements of the two Wilson loop operators in Fig.~\ref{fig:off_2d_pbc}. In this plot, we consider eigenstates within a thin energy shell $1.99<E_\alpha+E_\beta<2.01$ in all the momentum sectors up to (including) $k_x=2$ and $k_y=2$ on the $N_x=N_y=4$ lattice with $g^2=1$. The upper envelops of these distributions also exhibit bumpy plateaus at small and intermediate $\omega$ and show hints of an exponential fall-off at large $\omega$, although the statistics becomes marginal above $\omega=12.5$.

\subsection{Closed Boundary Condition}

In this subsection, we consider a triangular honeycomb lattice with a closed boundary condition, as shown in Fig.~\ref{fig:honeycomb_triangle}. The total number of plaquettes of a triangular lattice is fully determined by the number of plaquettes on each side $N$: $N_{\rm tot} = N(N+1)/2$. Closed boundary conditions assume that all links outside the boundary have vacuum quantum numbers $j=0$. With $\jmax=\frac{1}{2}$, the 2D SU(2) lattice gauge theory can be mapped onto a 2D spin model~\cite{Muller:2023nnk}
\begin{align}
\label{eq:H_closed}
H = & \sum_{(i,j)} \Big( h_+ \Pi^+_{i,j} - h_{++} \Pi^+_{i,j} \big( \Pi^+_{i+1,j} + \Pi^+_{i,j+1} + \Pi^+_{i+1,j-1} \big) \nn\\
& \quad\ \ + h_x (-0.5)^{c_{i,j}}\sigma_{i,j}^x \Big) \,,
\end{align}
where $h_+ = \frac{27\sqrt{3}}{8a^2g^2}$, $h_{++} = \frac{9\sqrt{3}}{8}g^2 $, and $h_x$ and $c_{ij}$ are the same as in the case of periodic boundary conditions. In the following, we will consider $g^2=1$. As explained earlier, any value of $g^2$ is expected to give qualitatively similar results for asymptotically large lattice sizes. 

As for the other systems investigated here, we exactly diagonalize the Hamiltonian and calculate matrix elements of certain operators in the energy eigenbasis. Here we will omit the studies of level statistics since the system has a discrete symmetry given by the dihedral group $D_3$ and focus on studying operators corresponding to the 1-plaquette and 2-plaquette Wilson loops, which were constructed in \cite{Muller:2023nnk}. Unlike the periodic case, here the results of operator matrix elements will depend the location of the operator. We consider two scenarios: Wilson loop operators defined at the center of the lattice, and those defined at a corner of the triangular lattice shown in Fig.~\ref{fig:honeycomb_triangle}.

\begin{figure}
\subfloat[Center.\label{fig:diag_2d_cbc_a}]{%
  \includegraphics[height=1.25in]{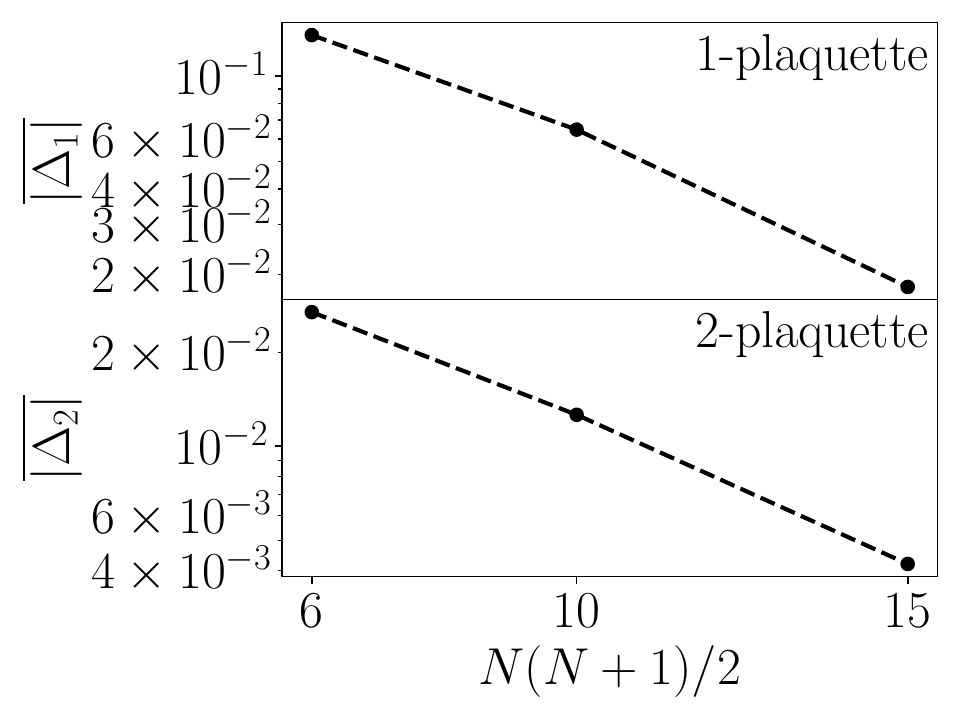}%
}\hfill
\subfloat[Corner.\label{fig:diag_2d_cbc_b}]{%
  \includegraphics[height=1.25in]{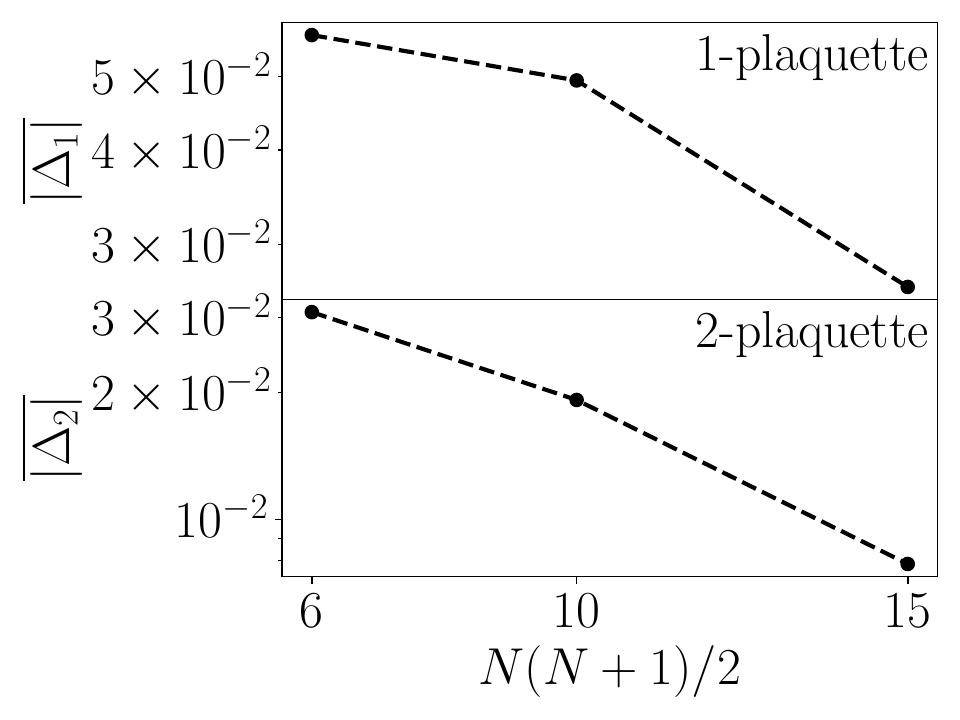}%
}
\caption{Averaged magnitude of the difference between the diagonal matrix element and the microcanonical ensemble average as a function of the lattice size, for Wilson loop operators defined at the center (left) and the left bottom corner (right).}
\label{fig:diag_2d_cbc}
\end{figure}

As in the periodic case we first plot the deviation of the diagonal matrix element from the microcanonical ensemble average in Fig.~\ref{fig:diag_2d_cbc}. For the operators defined at the lattice center, we clearly see an exponential decay of the averaged deviation magnitude as the system size increases (Fig.~\ref{fig:diag_2d_cbc_a}), which indicates the diagonal matrix elements are rapidly approaching the microcanonical expectation values. For the operators defined at the corner, we see a cusp in the 1-plaquette operator case (upper panel of  Fig.~\ref{fig:diag_2d_cbc_b}), which could be a boundary effect. However, for the 2-plaquette operator the exponential decrease is visible also for the corner location (lower panel of  Fig.~\ref{fig:diag_2d_cbc_b}).

\begin{figure}
\subfloat[Center.\label{fig:off_2d_cbc_a}]{%
  \includegraphics[height=1.25in]{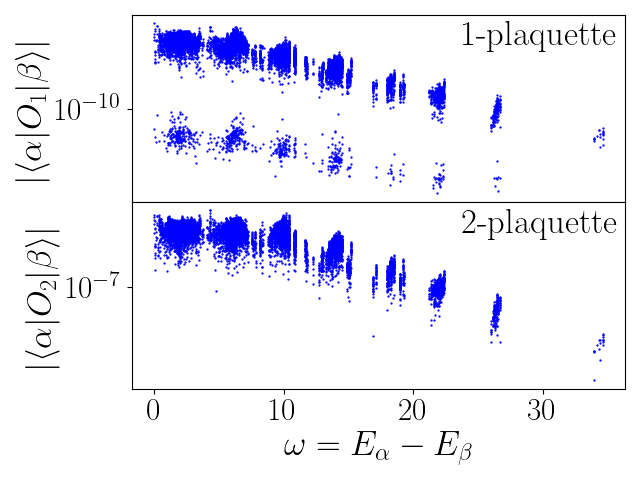}%
}\hfill
\subfloat[Corner.\label{fig:off_2d_cbc_b}]{%
  \includegraphics[height=1.25in]{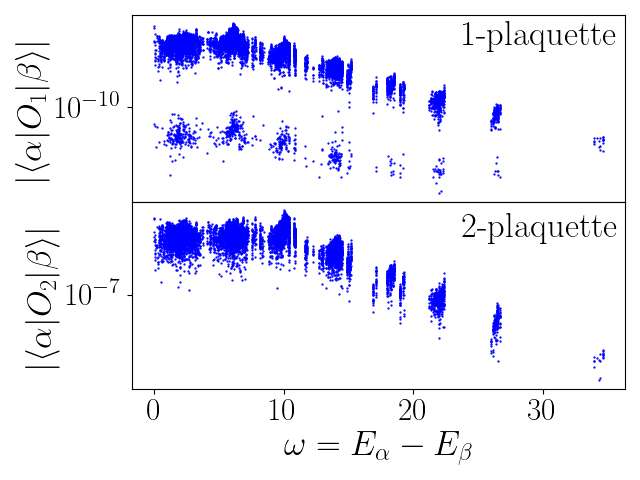}%
}
\caption{Magnitudes of off-diagonal matrix elements of the Wilson loop operators defined at the center (left) or left bottom corner (right). The two eigenstates are picked up from a narrow energy window $43.999<E_\alpha+E_\beta<44.001$, which leads to the discontinuity in $\omega$ here.}
\label{fig:off_2d_cbc}
\end{figure}

Then we study the off-diagonal matrix elements of the Wilson loop operators on the $N=5$ lattice. When calculating the matrix elements, we use only eigenstates within a thin energy shell $43.999<E_\alpha+E_\beta<44.001$. (One cannot simplify the Hamiltonian in the closed boundary case, i.e., Eq.~\eqref{eq:H_closed} by simply removing the constant terms, which shifts the zero energy reference such that the middle of the spectrum is around zero energy, as done in the case with a periodic boundary condition. This is because after removing the constant terms, one has to consider how the plaquettes (spins) just outside the boundary enters Eq.~\eqref{eq:H_closed}, which is different for the three edges of the triangular lattice. So we choose not to shift the zero energy reference here for convenience in the numerical construction of the Hamiltonian. This is why the values of the energy windows in the two cases with different boundary conditions look so different.) 

The results for the absolute values of the matrix elements are shown in Fig.~\ref{fig:off_2d_cbc} as a function of $\omega = E_\alpha-E_\beta$. Again, we notice a bumpy plateau in the small $\omega$ region ($\omega\lesssim 10$) and an exponential fall-off at large $\omega$ for both operators, no matter whether they are defined at the center or corner. In the case of the 1-plaquette operator, some off-diagonal matrix elements have much smaller values ($\lesssim10^{-13}$) than others, which are essentially zero within the numerical precision. This is a result of discrete symmetries in the system. The 1-plaquette operator defined at the center is symmetric under the dihedral group $D_3$ on the $N=5$ triangular lattice. As a result, its matrix element between two eigenstates with different symmetry under $D_3$ transformations vanishes by virtues of the Wigner-Eckart theorem. Similarly, the 1-plaquette operator defined at a corner of the triangular lattice respects the reflection around the perpendicular bisector line going through this corner. Thus its matrix element between two eigenstates that have different reflection symmetry vanishes. We do not see vanishing off-diagonal matrix elements in the case of 2-plaquette operators, since they break the $D_3$ symmetry of the lattice.

\begin{figure}
\subfloat[$0<\omega<4$.\label{fig:gauss_fit_off_O2_a}]{%
  \includegraphics[height=1.25in]{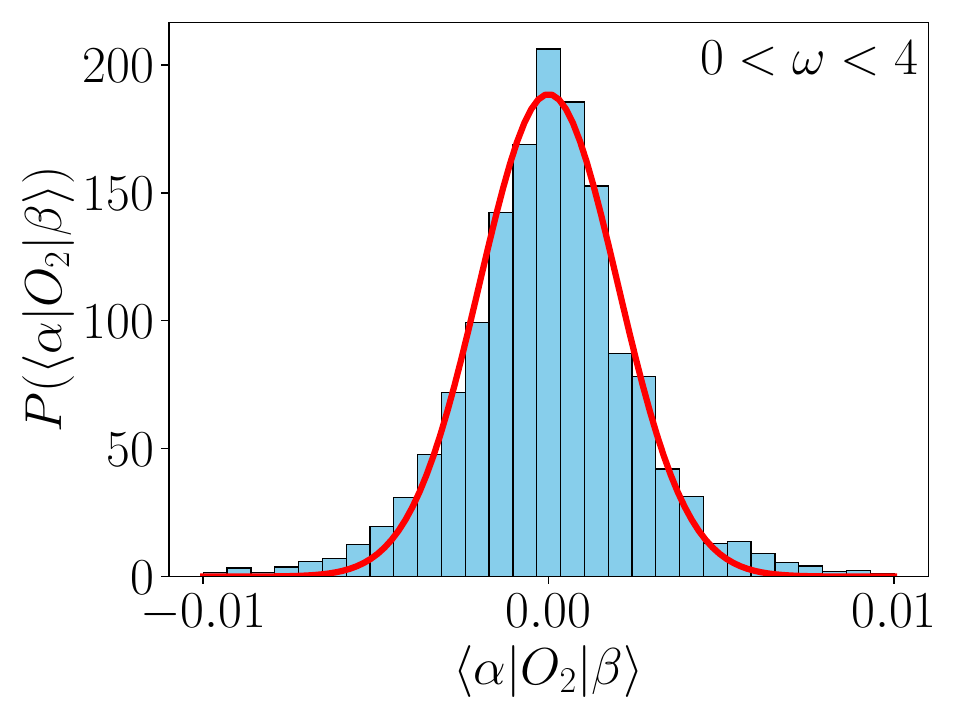}%
}\hfill
\subfloat[$4<\omega<8$.\label{fig:gauss_fit_off_O2_b}]{%
  \includegraphics[height=1.25in]{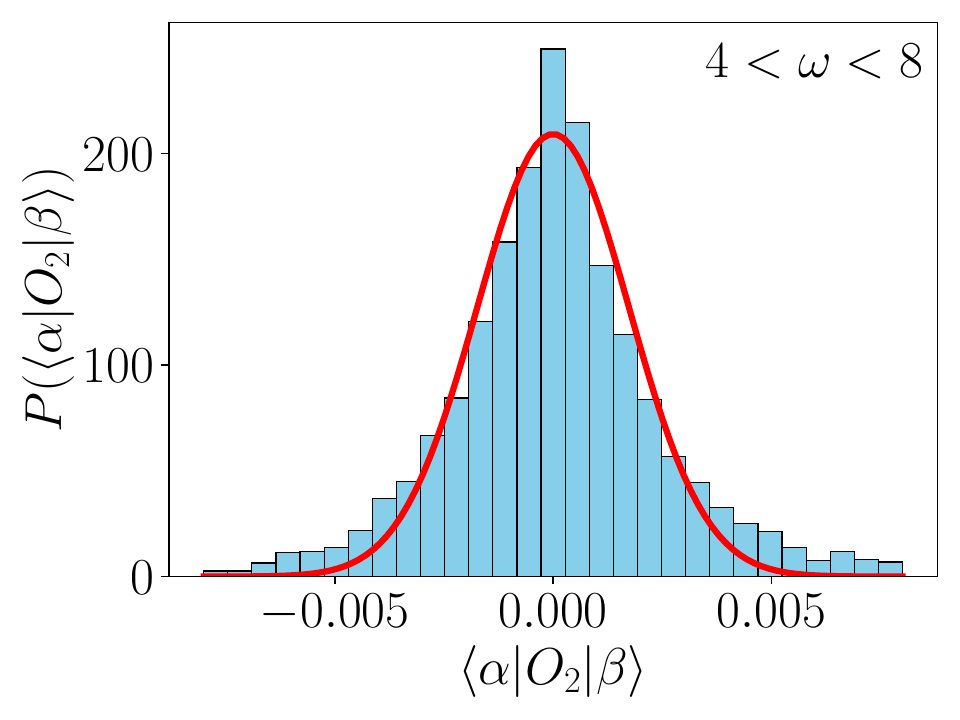}%
}
\caption{Gaussian fits for distributions of $\langle n |O_2 | m \rangle$ (the operator is defined at the center) in two $\omega$ slots: $0<\omega<4$ (left) and $4<\omega<8$ (right). The fitted Gaussian widths are roughly $0.00200$ and $0.00174$ for the left and right panel, respectively.}
\label{fig:gauss_fit_off_O2}
\end{figure}

Finally, we study whether the off-diagonal matrix elements of the Wilson loop operators can be well described by Gaussian distributions. To this end, we consider the 2-plaquette operator defined at the center of the $N=5$ lattice studied above and use the eigenstates contained in the same energy shell. The distributions of the off-diagonal matrix elements $\langle n| O_2 |m\rangle$ in different $\omega$ regions are plotted in Fig.~\ref{fig:gauss_fit_off_O2} together with Gaussian fits. The two different $\omega$ windows are $0<\omega<4$ and $4<\omega<8$. We see in the smaller $\omega$ region, the distribution of $\langle n |O_2 | m \rangle$ can be better described by a Gaussian, which is consistent with the results obtained for the plaquette chain and the honeycomb lattice with periodic boundary conditions.

\section{Cutoff Convergence}
\label{sec:Convergence}

As explained in Section~\ref{sec:Intro} we are interested in the double limit of large number of plaquettes $N$ and large enough $\jmax$ to reach convergence. This is not feasible for us with our present computer resources. Therefore we investigate separately the cases $N$ large, $\jmax=\frac{1}{2}$ (Section~\ref{sec:Ising_chain}  and \ref{sec:Ising_honeycomb}) and the case $N=3$, $\jmax\leq \frac{7}{2}$ (this section). More precisely we investigate the validity of the ETH relation Eq.~(\ref{eq:ETH}) for a linear chain of three plaquettes with periodic boundary conditions in an energy region where we observe good convergence with the cutoff $\jmax$, i.e., we obtain results for the KS Hamiltonian in a converged Hilbert space region. 

\begin{figure}[t!]
	\centering
	\includegraphics[width=0.95\linewidth]{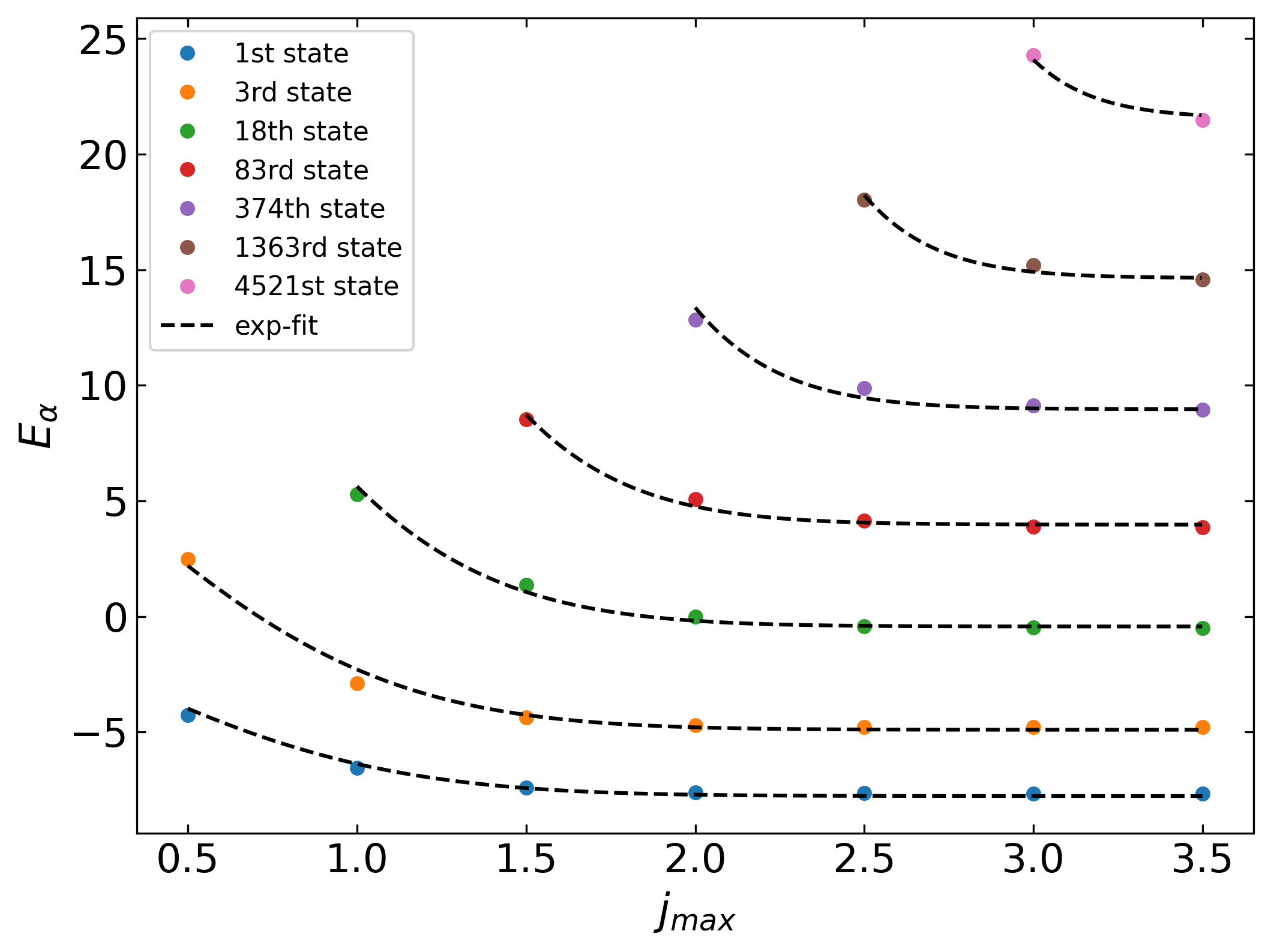}
	\caption{Some energy eigenvalues for the $k=0$, positive parity and positive top-down symmetry sector plotted against the cutoff for $g^2=0.8$ with a corresponding exponential fit of the form of Eq.~\eqref{eq:conv_beh}.}
	\label{fig:eigenvalues_convergence}
\end{figure}

We start by examining this convergence behavior. A few typical examples of eigenenergies for $g^2=0.8$ are shown in Fig.~\ref{fig:eigenvalues_convergence}. Eigenstates with different momentum $k$ (which is discretized) and/or parity do not mix. Therefore, to study the characteristic RMT behavior in the eigenenergy spectrum one has to choose a specific sector. We concentrate on the sector with momentum $k=0$ and positive parity. An additional symmetry, namely top-bottom symmetry, arises for $\jmax > \frac{1}{2}$. In this case we chose the sector corresponding to the eigenvalue $+1$. 
Obviously, very good convergence can be achieved for states that are not too highly excited, i.e., $E-E_1 \lesssim 20$, where $E_1$ denotes the ground state energy of the system. This convergence can be tracked quantitatively by fitting the eigenvalues as function of $j$ by simple exponential curves of the form 
\be
\label{eq:conv_beh}
E_{\alpha} (j)=A_{\alpha}e^{- j(j+1)g^{2}}+\lambda_{\alpha} \, ,
\ee
and extrapolating to infinite $j$ (see dashed lines in Fig.~\ref{fig:eigenvalues_convergence}). Here $\lambda_{\alpha}$ is the $\alpha$-th energy eigenvalue in the limit $j \rightarrow \infty$ and $A_{\alpha}$ is a fit parameter which does not depend on $\jmax$ but on $\alpha$. We observe a power law behavior of the parameter $A_{\alpha}$ with respect to $\alpha$ which we use to further improve the extrapolation: We fit $A_{\alpha}$ by a three-parameter function $A_{\alpha} = a \cdot \alpha^{b}+c$. This finally leaves a function $E_{\alpha} (j)$ with only one parameter, namely $\lambda_{\alpha}$, allowing us to define a precise criterion for a converged eigenvalue, 
\be
\label{eq:conv_criteria}
\frac{E_{\alpha} (\jmax)-\lambda_{\alpha}}{\lambda_{\alpha}} < 5\% \, ,
\ee
where in our case $\jmax$ is $\frac{7}{2}$. Using this criterion we find convergence for eigenvalues up to $E_{\rm max}=24.3$. It is clear from Fig.~\ref{fig:eigenvalues_convergence} that $\jmax=\frac{1}{2}$ is typically far from convergence at this value of the coupling constant ($g^2 = 0.8$), which is not surprising since even the ground state energy becomes exact for $\jmax = \frac{1}{2}$ only in the strong coupling limit. 

In the selected converged energy regions we proceed with the analysis of ETH properties. In line with our discussion in Section~\ref{sec:Intro} we choose a ``physical'' observable, namely the electric energy $H_{\rm el}$, in a specific symmetry sector to check the validity of Eq.~(\ref{eq:ETH}). Furthermore, we study the structure of the spectral response function $f(E,\omega)$ in more detail. Since the possibilities of creating gauge invariant states on our lattice increase with increasing $\jmax$, new energy eigenvalues emerge such that the available statistics improve rapidly with increasing $\jmax$.  

\begin{figure}[t!]
	\centering
	\includegraphics[width=0.95\linewidth]{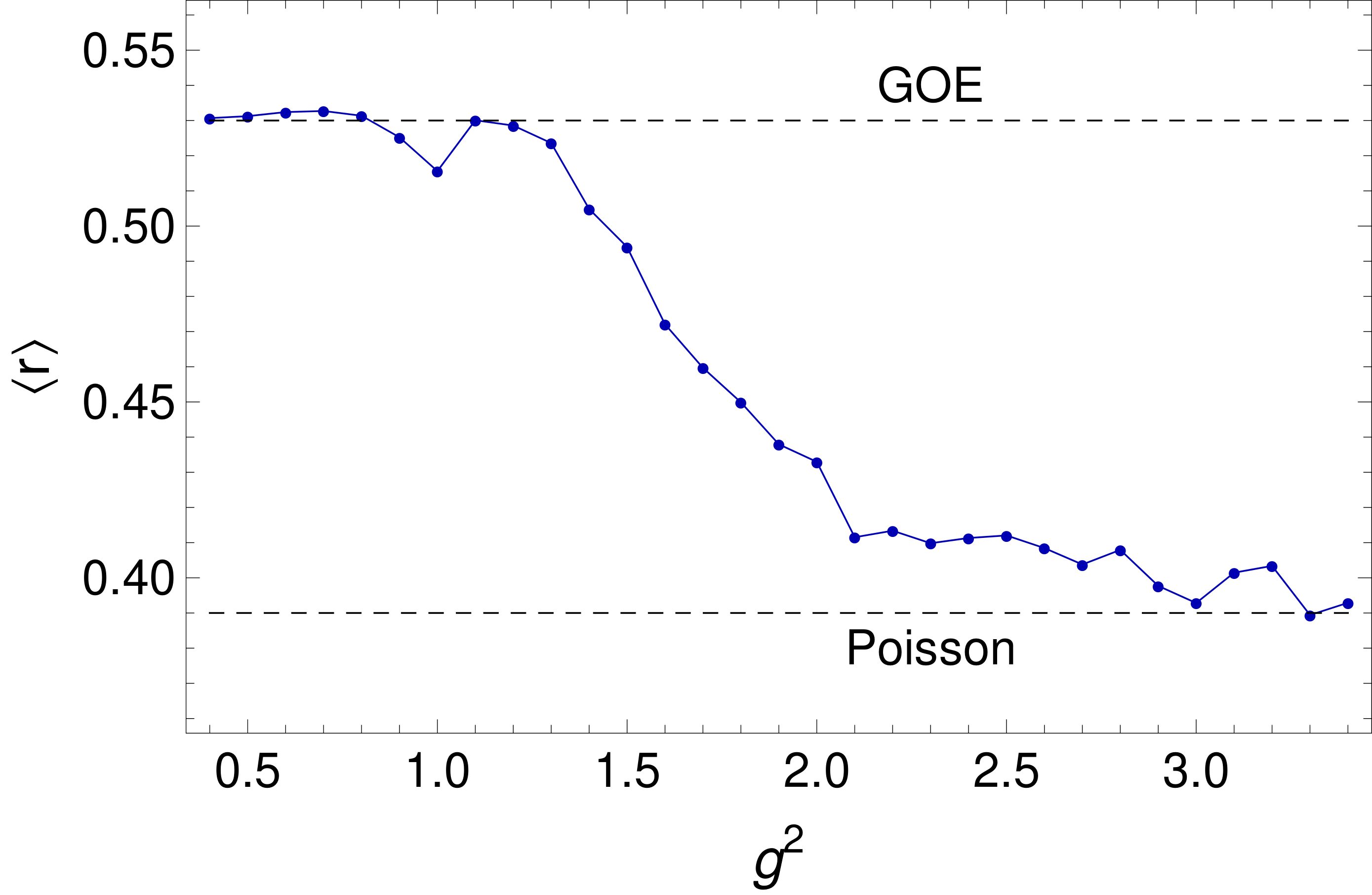}
	\caption{Average restricted gap ratio $\langle r \rangle$ plotted against the coupling constant $g^2$ for $N=3$ plaquettes and $\jmax=3.5$. The value for a GOE distribution is 0.53 and that for a Poisson distribution is 0.39. Obviously $\langle r\rangle$ extrapolates between the ergodic and non-ergodic regime as a function of $g^2$.}
	\label{fig:Restricetd gap ratio}
\end{figure}

Again, we proceed as in Sections \ref{sec:Ising_chain} and \ref{sec:Ising_honeycomb} and follow the template of similar investigations of possible ETH behavior (see e.g.\ \cite{Jansen2019ETH}): We first check for RMT behavior in the eigenenergy spectrum, before we analyze the diagonal, and then the off-diagonal matrix elements of Eq.~(\ref{eq:ETH}). However, the KS Hamiltonian has a special property one needs to be aware of. As one can see in Eq.~(\ref{eq:KS_Hamiltonian}) the KS Hamiltonian has an electric and a magnetic part. A sufficient contribution of the magnetic part to the overall Hamiltonian is essential for observing chaotic behavior. Thus, for fixed lattice spacing $a$ and size $N$ the degree of ergodicity depends on $g$. For sufficiently large $g$ the theory becomes non-ergodic and the resulting energy level statistics approach a Poisson distribution while for moderate $g$ its RMT behavior should be clearly visible. For example the mean gap ratio should interpolate smoothly between the Poisson and RMT (GOE) values, which is exactly what we observe in Fig.~\ref{fig:Restricetd gap ratio}. This plot also shows in which range $g^2$ has to be chosen to observe ergodic properties for $N=3$, namely $g^2\leq 1.2$. We choose typically $g^2=0.8$ or $g^2=1$. Smaller values of $g^2$ are not practical, because the slow convergence with respect to $\jmax$ would yield an insufficient converged region size.

\begin{figure}[t!]
	\centering
	\includegraphics[width=0.95\linewidth]{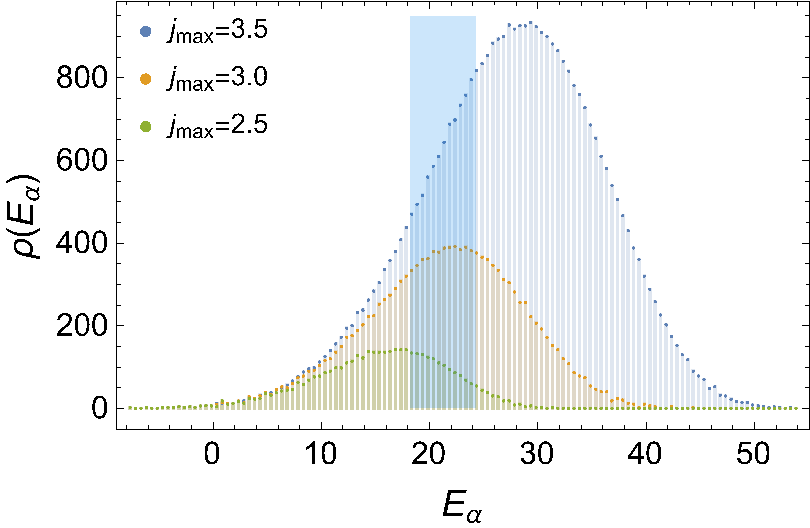}
	\caption{Histograms of the whole energy spectra for $\jmax\in\{\frac{5}{2},3,\frac{7}{2}\}$ and $g^2=0.8$. The blue rectangle depicts the converged region, leaving out states affected by finite size effects.}
	\label{fig:hist_different_jmax}
\end{figure}

Overall, requiring convergence with respect to $\jmax$, a value of $g^2$ for which the system behaves ergodically, and the magnitude of discretization and finite size artifacts strongly limit the usable statistics and parameter range. This is illustrated in Fig.~\ref{fig:hist_different_jmax} for the histograms of the complete spectra for electric field cutoffs $\jmax\in\{\frac{5}{2},3,\frac{7}{2}\}$.  While there is still a large difference in the level density in a wide energy region between $\jmax = 3$ and $\jmax = \frac{7}{2}$, our convergence studies described above allow us to assess the convergence beyond $\jmax = \frac{7}{2}$, i.e.,  confirm that $\rho_{7/2}(E) \approx \rho_\infty(E)$ for $E\lesssim E_{\rm max}$, where the index of $\rho_j(E)$ denotes the $j$-cutoff.
In order to suppress finite size effects, we discard 14\% of the states at both edges of the full spectrum. As a result, the low-energy part of the converged spectrum is not taken into account. The truncated converged energy window, used in our analysis, is thus determined by the upper bound derived from our convergence criterion (\ref{eq:conv_criteria}) and the lower bound from the finite size truncation. The converged window is highlighted by a blue shaded rectangle in Fig.~\ref{fig:hist_different_jmax}. The remaining number of states and matrix elements of  $H_{\rm el}$ is large enough to obtain rather precise results.

\begin{figure}[t!]
	\centering
	\includegraphics[width=0.95\linewidth]{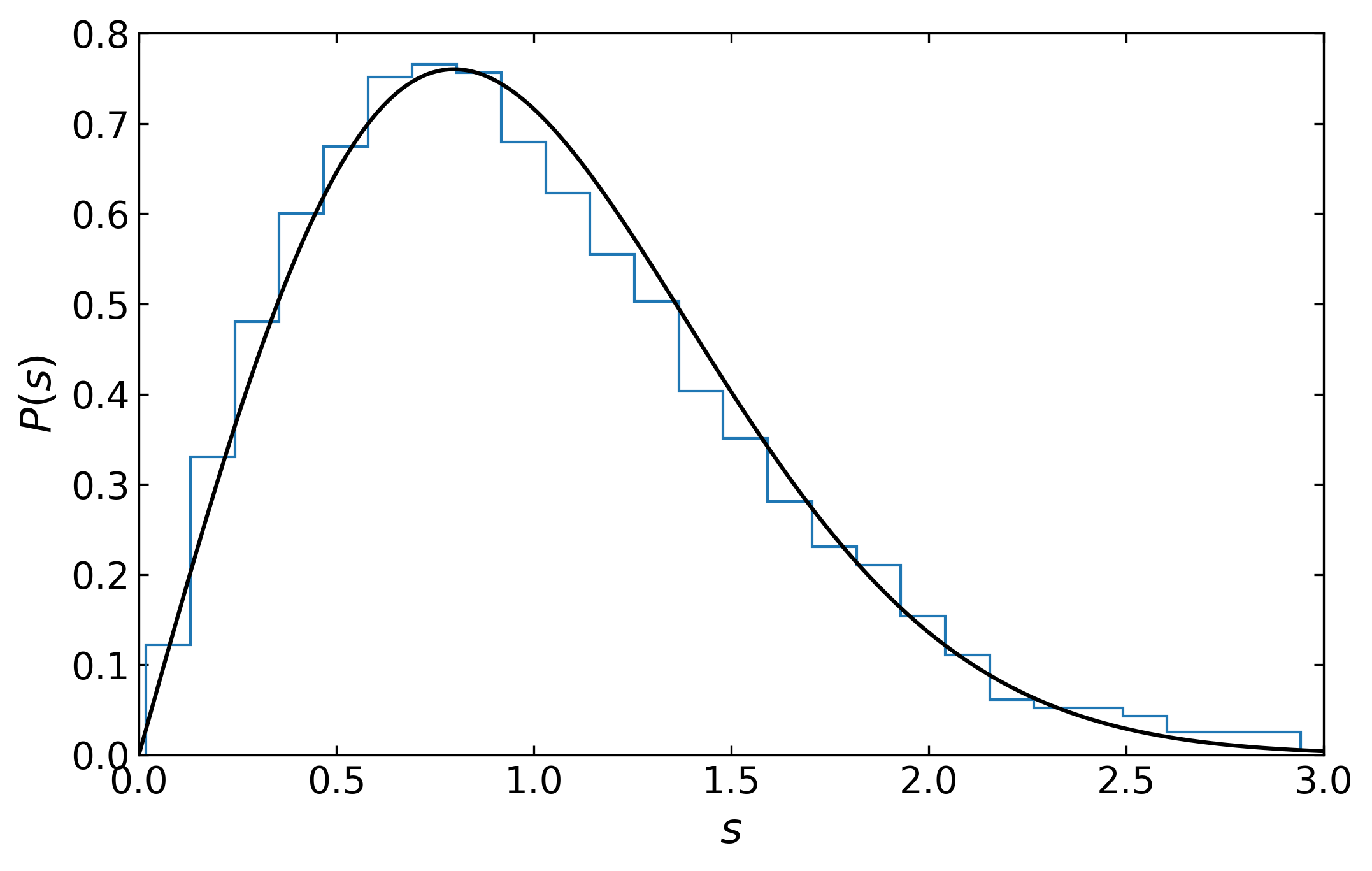}
\caption{The normalized level spacing $s=\delta_\alpha/\bar{\delta}$ (as defined in Section~\ref{subsec:Periodic Boundary Condition}) distribution in the converged energy window compared to the Wigner-Dyson distribution $P(s)$.}
	\label{fig:GOE conv region}
\end{figure}

Another standard test for GOE behavior in the eigenenergy spectrum is the nearest-neighbor level spacing statistics. Figure~\ref{fig:GOE conv region} shows a histogram of the normalized level spacing distribution $P(s)$ in the converged energy window for $g^2=0.8$ in comparison with the Wigner-Dyson distribution \eqref{eq:Wigner_Dyson_distribution}. The agreement is good, but not perfect, suggesting that there are still deviations from exact RMT behavior in this energy window.

\begin{figure}[t!]
	\centering
	\includegraphics[width=0.95\linewidth]{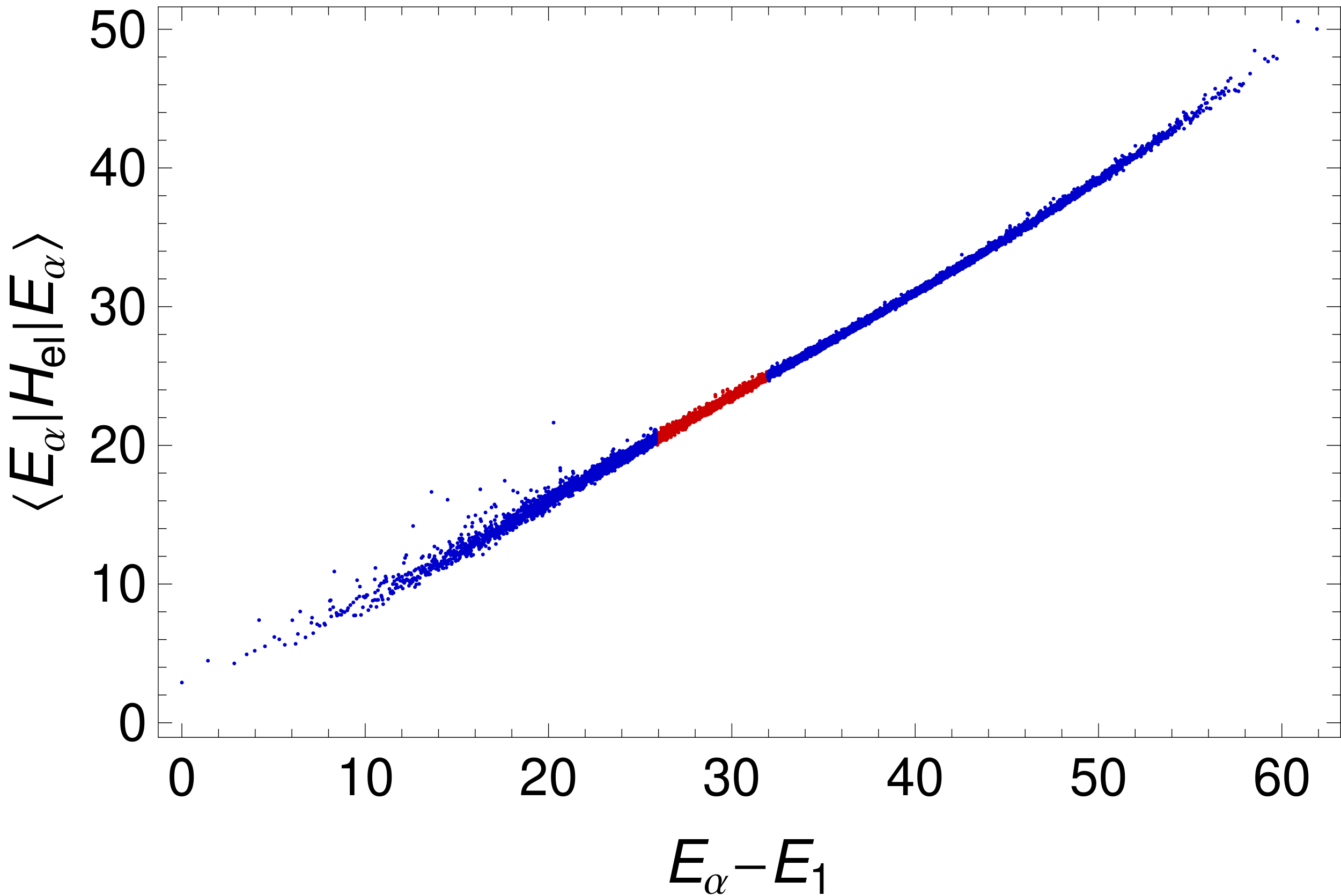}
	\caption{Diagonal matrix elements of the electric energy in the energy basis $\langle E_{\alpha}|H_{el}|E_{\alpha}\rangle$ plotted against the the respective energy eigenvalues $E_{\alpha}-E_{1}$. The red colored part corresponds to the converged energy window used}
	\label{fig:diagonal_ME}
\end{figure}

Next, we analyze the diagonal part of Eq.~(\ref{eq:ETH}) for the matrix elements of the electrical energy operator in the specified symmetry sector. Equation~(\ref{eq:ETH}) predicts the presence of exponentially suppressed rapidly varying fluctuations superimposed on a smooth function of $E$. This ensures the existence of a well defined microcanonical ensemble average that is obtained by averaging $\langle E_{\alpha}|H_{\rm el}|E_{\alpha} \rangle$ over a small energy window. Figure~\ref{fig:diagonal_ME} shows the distribution of the expectation value of the electric energy as a function of the energy for all energy eigenstates of the system in the selected symmetry sector. The values for states inside the chosen energy window are highlighted in red. The spread of the distribution around a nearly linear dependence on energy is seen to be small, i.e., the electric energy operator has a well defined microcanonical average $\langle H_{\rm el}\rangle_{\rm mc}(E)$ over a wide range of the spectrum. The width of the distribution is expected to shrink exponentially with the lattice size as observed for the plaquette operators in the truncated ($\jmax = \frac{1}{2}$) version of the gauge theory (see Figs.~\ref{fig:diag_2d_pbc} and \ref{fig:diag_2d_cbc}).

As an aside let us note that Fig.~\ref{fig:diagonal_ME} shows that the electrical energy contributes the dominant part to the total energy, with $E_{\rm el}$ ranging between $0$ and $56.7$, whereas the magnetic energy is limited to the range $-15 < E_{\rm mag} < 15$ in lattice units. This shows that the system is still far from the thermodynamic limit of the continuum gauge theory.

\begin{figure}
\subfloat[\label{fig:offdiagonal_convmean}]{%
  \includegraphics[width=0.48\linewidth]{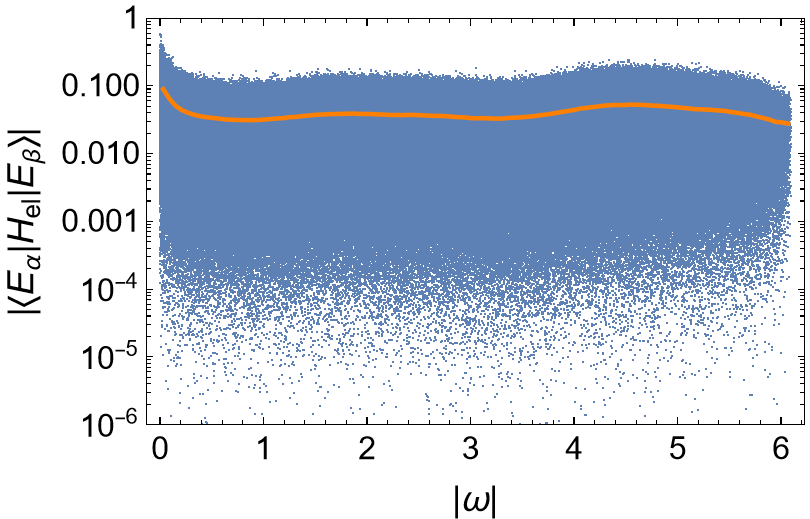}%
}\hfill
\subfloat[\label{fig:mvav_convmean2}]{%
  \includegraphics[width=0.48\linewidth]{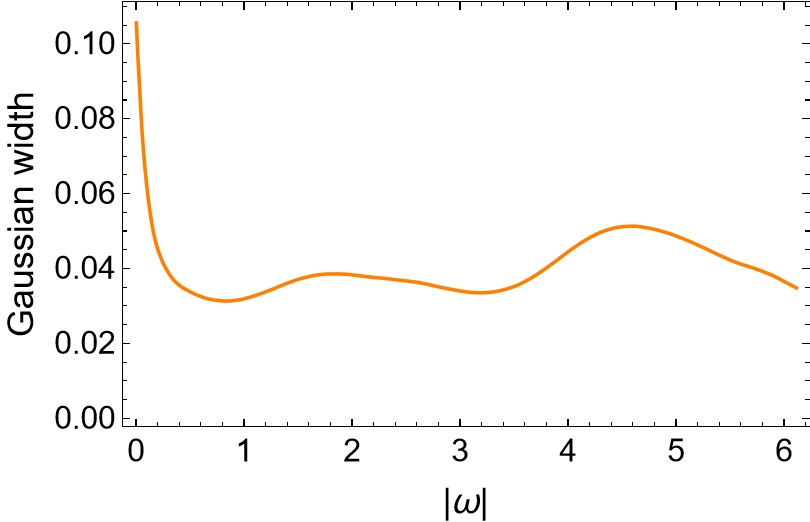}%
}
\caption{Absolute values of the off-diagonal matrix elements of the electric energy in the energy basis $|\langle E_{\alpha}|H_{\rm el}|E_{\beta}\rangle|$ against $\omega = E_\alpha-E_\beta$ in the energy window $|(E_\alpha + E_\beta)/2 - \bar{E}| < \Delta E/2$ with $\bar{E}=21.5$ and $\Delta E=1$. Part (a) presents a logarithmic plot of all matrix elements (blue) and their Gaussian widths (orange), obtained by the second moment method, whereas (b) shows a linear plot of the Gaussian widths only.
}
\label{fig:diag_mvaverage_cutoff}
\end{figure}

We now proceed with the analysis of the off-diagonal matrix elements of the electric energy, i.e., $\langle E_{\alpha}|H_{\rm el}|E_{\beta}\rangle$, in the converged energy window. Figure~\ref{fig:diag_mvaverage_cutoff} shows these matrix elements in the energy window 
\begin{equation}
|(E_\alpha + E_\beta)/2 - \bar{E}| < \Delta E/2
\end{equation}
with $\bar{E}=21.5$ and $\Delta E=1$ for $g^2=0.8$. In order to analyze the spectral behavior with respect to $|\omega| = |E_\alpha-E_\beta|$ we calculate the second moment as defined in Eq.~(\ref{eq:second moment}).

For the spectral function one expects several characteristic $\omega$ regions (see Alessio {\it et al.} \cite{DAlessio:2015qtq}, Section 4.3.1.2): an exponential decay at large $\omega$, a bumpy region at intermediate $\omega$ reflecting quasiparticle contributions, and a diffusive plateau at the smallest values of $\omega$. For nonabelian gauge theories an effective description in terms of quasiparticles is typically based on analogs of glueballs; at high excitation energy such quasiparticles may or may not exist depending on the strength of the coupling. Such quasiparticles show up as broad peaks in the spectral function. Typically these structures are quite broad but details are specific to the theory under investigation. These properties are visible in Figs.~\ref{fig:offdiagonal_convmean} and \ref{fig:mvav_convmean2}. 

\begin{figure}[ht]
	\centering
	\includegraphics[width=0.95\linewidth]{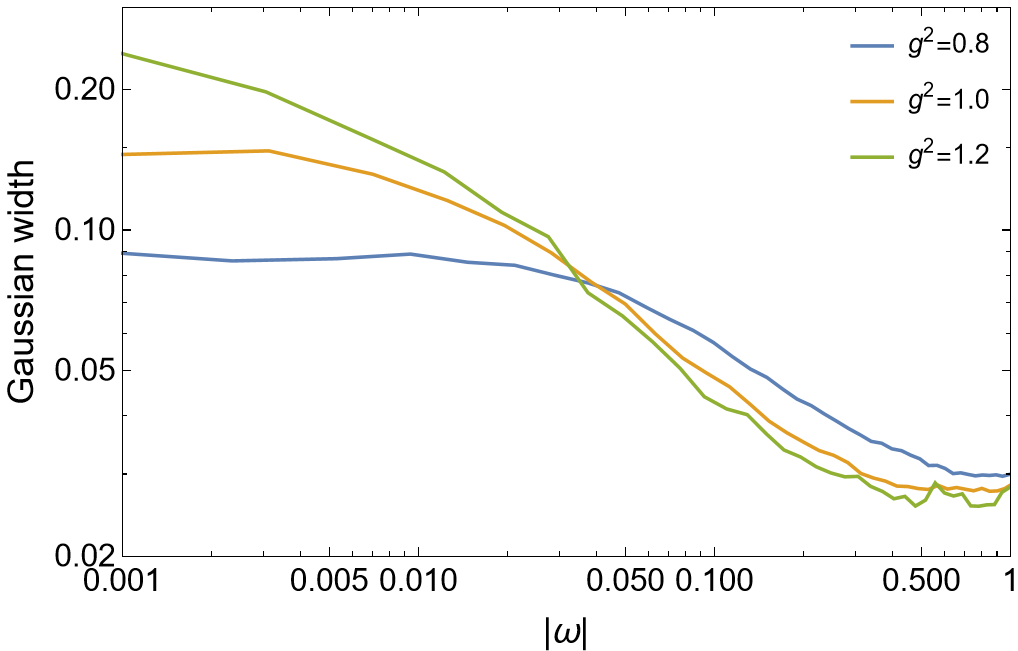}
	\caption{Gaussian widths of off-diagonal matrix elements, obtained by the second moment method, for the equivalent energy windows $\bar{E}=E_{1}+31$ and $\Delta E=1$ for coupling constant $g^2\in \{0.8,1.0,1.2\}$ in the small $\omega$ region.}
	\label{fig:diffusive_peak}
\end{figure}

For many systems one also observes a transport peak and a diffusive plateau at very low $\omega$. Whether a prominent transport peak exists for nonabelian gauge theories and, if so, for which operators and in which parameter range, is still under investigation \cite{Akamatsu:2011mw,Brandt:2015aqk,Casalderrey-Solana:2018rle}.  Interestingly, our results shown in Figs.~\ref{fig:diag_mvaverage_cutoff} and \ref{fig:diffusive_peak} support the existence of such a diffusive transport peak, at least for the pure SU(2) gauge theory.

\begin{figure}[ht]
	\centering
	\includegraphics[width=0.95\linewidth]{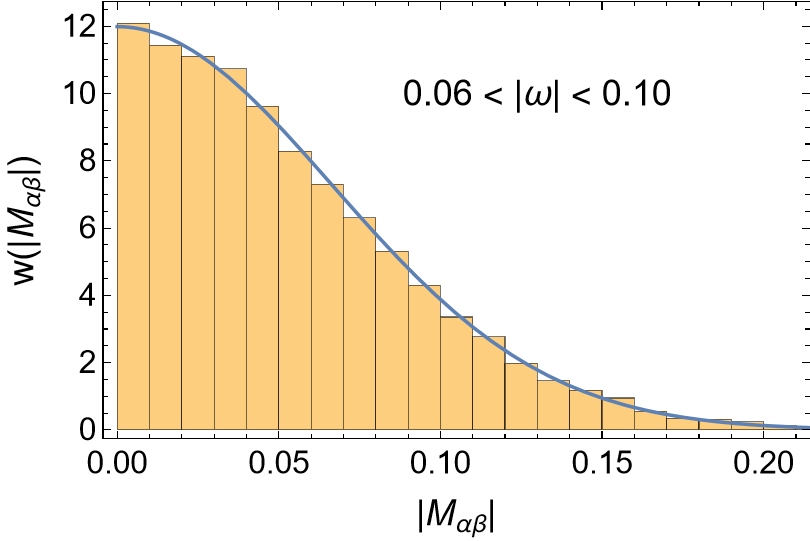}
	\caption{Distribution of matrix elements $|M_{\alpha\beta}|=|\langle E_{\alpha}|H_{\rm el}|E_{\beta}\rangle|$ of the total electric energy operator $H_{\rm el}$ in the range $0.06 < |\omega| < 0.10$ for $g^2=0.8$, shown together with a Gaussian fit.}
	\label{fig:gauss_slice}
\end{figure}

\begin{figure}[ht]
	\centering
	\includegraphics[width=0.95\linewidth]{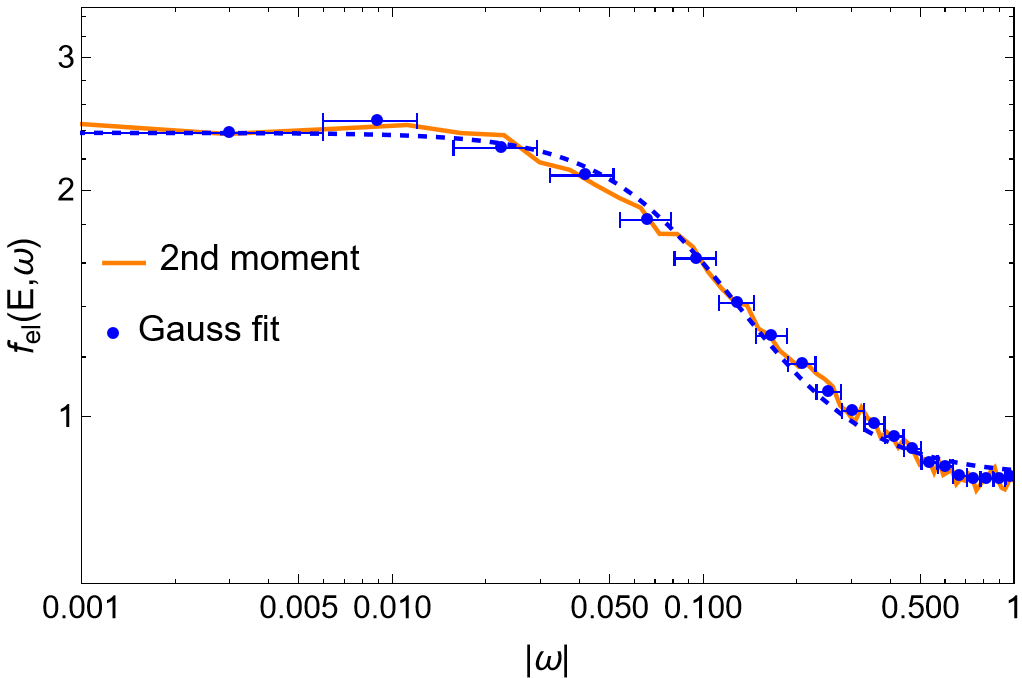}
	\caption{Double logarithmic plot of the spectral function $f_{\rm el}(E,\omega)$, which is obtained by both the second moment and Gaussian fit methods in the energy window defined by $\bar{E}=23.5$ and $\Delta E=1$. The blue bars represent the $\omega$ regions of matrix elements used for the Gaussian distribution fit. The blue dashed line represents an analytical fit of the blue dots, which is of the form $a/(\omega^2+b^2) + c$ for small $\omega$. We find $a\approx1.50\times10^{-2}$, $b\approx9.81\times10^{-2}$ and $c\approx0.835$.}
	\label{fig:spect_fct_cutoff}
\end{figure}

\begin{figure}[ht]
	\centering
	\includegraphics[width=0.95\linewidth]{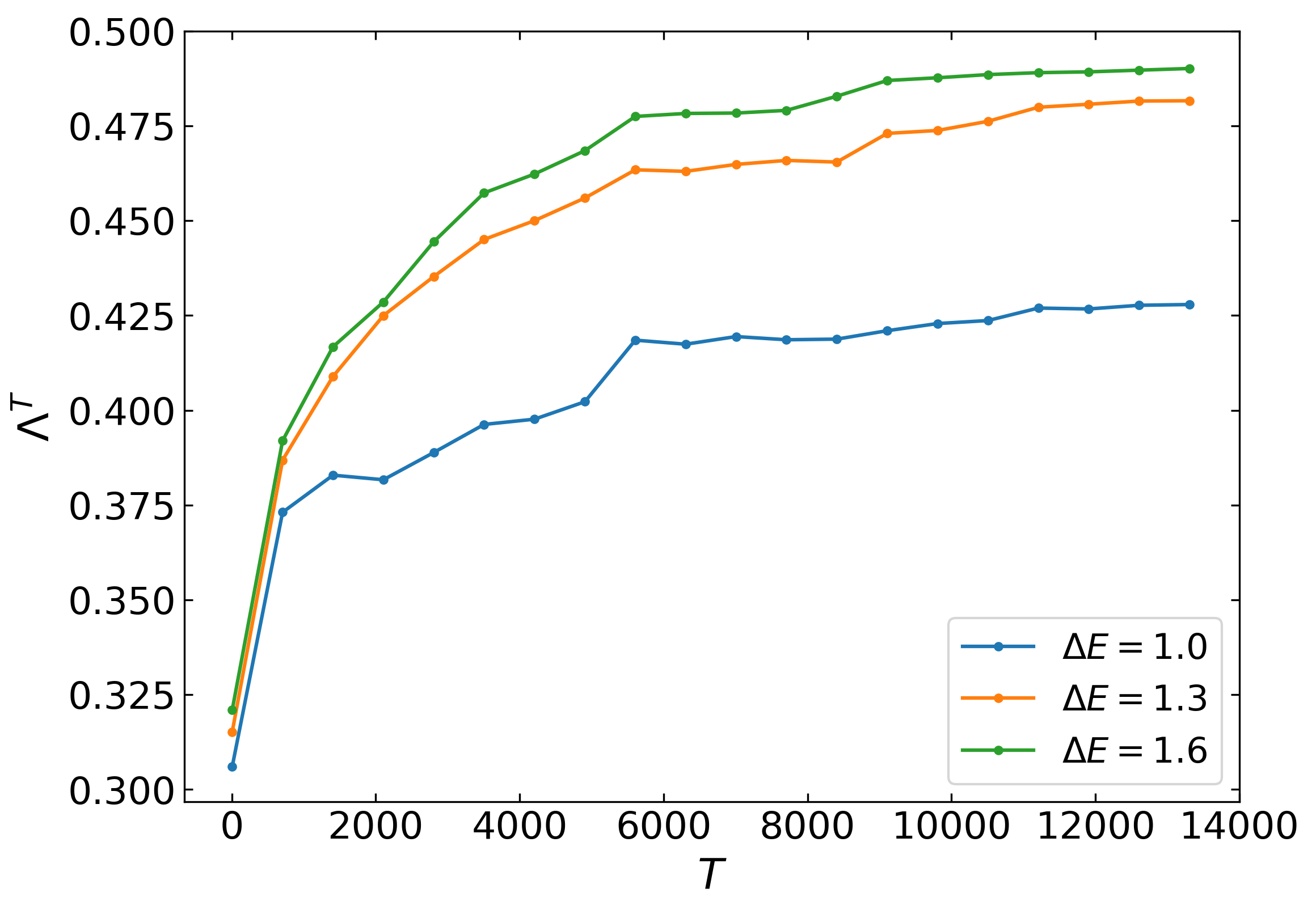}
	\caption{$\Lambda^T$ measure for the electric energy within the energy windows defined by the mean energy $\bar{E}=23.5$ and the window sizes $\Delta E=1$, $\Delta E=1.3$ and $\Delta E=1.6$ for $g^2=0.8$.}
	\label{fig:Lambda_T_measure_conv}
\end{figure}

Therefore we now focus our analysis specifically on the small $\omega$ regime, i.e., the transport peak region.
We start by discussing the $g$ dependence of the transport peak, shown in Fig.~\ref{fig:diffusive_peak}. In order to be able to compare the behavior for different coupling constants we define the equivalent energy regions with respect to each ground state. In the following we restrict ourselves to the energy window with $\bar{E}-E_{1}(g^2)=31$ and $\Delta E=1$.
We know from our discussion of Fig.~\ref{fig:Restricetd gap ratio} that the chaotic nature of our system is lost if the coupling constant exceeds $g^2\approx 1.2$. As the diffusive transport peak is part of this dynamics, it is natural to expect it to become less well defined when $g^2$ approaches or exceeds 1.2. Indeed, Fig.~\ref{fig:diffusive_peak} suggests that the diffusive plateau at small $|\omega|$ disappears right around this value of the coupling constant although there is still a pronounced peak in the spectral function. 

As explained in Section \ref{sec:Ising_honeycomb} the spectral function can be obtained in different ways. On one hand it can be defined via the second moment of matrix elements as already shown in Figs.~\ref{fig:diag_mvaverage_cutoff} and \ref{fig:diffusive_peak}. On the other hand, the distribution of matrix elements magnitudes in small $\omega$ windows appears Gaussian, suggested by the central limit theorem, which is clearly supported by Fig.~\ref{fig:gauss_slice}. Similar plots are obtained in the whole $\omega$ spectrum. Thus we are in the position to define the spectral function from Eq.~(\ref{eq:M^2}), using the Gaussian width of the matrix element distributions.
The comparison of these two methods for $g^2=0.8$ is depicted in Fig.~\ref{fig:spect_fct_cutoff}. Both the second moment and the Gaussian fit methods show very good agreement, even for the diffusive plateau $|\omega|<0.02$, supporting that the off-diagonal matrix element magnitudes follow a Gaussian distribution. The plot also supports a functional form of a diffusive transport peak, i.e., Eq.~(\ref{eq:peak_fit}).

Finally we want to discuss certain limitations of the GOE indicator used by considering the $\Lambda^T$ measure defined in Eq.~(\ref{eqn:lambdaT}). Unlike the above analyses, the calculation of the $\Lambda^T$ measure does include the original matrix elements without taking the absolute value. In Fig.~\ref{fig:Lambda_T_measure_conv} we show $\Lambda^T$ for three different sizes of the energy windows, namely $\Delta E=1$, $\Delta E=1.3$ and $\Delta E=1.6$, around the same mean energy $\bar{E}=23.5$ (all of which lie in the blue band denoting the converged energy region indicated in Fig.\ref{fig:hist_different_jmax}). All three plots exhibit the expected monotonic growth with increasing $T$, i.e., decreasing number of allowed nonzero matrix elements, and convergence to some saturation value $\Lambda^\infty_{\Delta E}$. In the case of $\Delta E=1$ at $T=14,000$, we find this value to be $\Lambda^T_{1}\approx0.429$, where we are left with 905 nonzero matrix elements. For the two larger energy windows and the same $T$ we obtain $\Lambda^T_{1.3}\approx0.482$ and $\Lambda^T_{1.6}\approx0.490$ with 1182 and 1474 nonzero matrix elements, respectively. We observe a sustained approach to the GOE prediction $\Lambda_{\rm GOE}=0.5$ for increasing energy window width $\Delta E$. We interpret this as a result of the larger number of nonzero matrix elements at each $T$ as the energy window width $\Delta E$ increases. This finding highlights our numerical limitations for small energy windows, e.g., the case $\Delta E=1$. These results also show similar saturation time scales of $\Lambda^T$ for the electric energy $H_{\rm el}$ of order $T \sim 10^4$ as found in Section~\ref{sec:Ising_honeycomb}.

\section{Summary and Outlook}
\label{sect:conclusions}

We investigated several truncated versions of lattice-discretized 2+1 dimensional SU(2) gauge theory with respect to the properties underpinning the Eigenstate Thermalization Hypothesis. The systems we studied include linear plaquette chains up to length $N=19$, honeycomb lattices with up to 20 plaquettes with the electric field Hilbert space truncated at $\jmax = \frac{1}{2}$, and the $N=3$ plaquette chain on the fully converged electric field Hilbert space. Our results are encouraging: All these truncated versions of lattice gauge theory exhibit clear signs of behavior that is consistent with the ETH: The energy level spectrum obeys GOE statistics within the limits of statistical and systematic uncertainties. The fluctuations of diagonal matrix elements of the operators we studied (the total electric energy as well as 1- and 2-plaquette Wilson loops) were found to exponentially decay with the lattice area. We studied the off-diagonal matrix elements of several operators between nearby energy eigenstates and found their magnitudes follow Gaussian distributions. Their signs are correlated and only become more random in smaller $\omega$ windows. We also calculated the spectral function $f(E,\omega)$ for all three truncated LGT models and found that they conform to the expectations for a system that exhibits quantum chaos.

While encouraging, these pioneering studies were limited to small systems due to lack of computer resources and thus rather constitute a proof of principle than definitive evidence that 2+1 dimensional SU(2) lattice gauge theory exhibits ETH behavior in the continuum limit. However, they motivate hope that along these lines the numerical evidence can be significantly strengthened if more computer resources are invested.  In addition to substantial hardware resources this will also require the development and implementation of efficient algorithms for the calculation of matrix elements between energy eigenstates in very large Hilbert spaces, such as the kernel polynomial method \cite{Weisse:2006zz}. 

The following bullet points illustrate the range of possible future work extending the results reported here:
\begin{itemize}
\item 
Explore the convergence of Hilbert space truncations for lattices with a larger number of plaquettes.
\item 
Extend our investigations to more operators ${\cal A}$ to better understand for which operators the defining ETH relation (\ref{eq:ETH}) holds. 
\item 
Extend our studies to three spatial dimensions using point-splitting methods as in \cite{Zache:2023dko}.
\item 
Determine the Thouless energy for 2+1 dimensional SU(2) gauge theory if it exists, i.e., the value of $\Delta E$ below which the RMT behavior in the off-diagonal matrix elements is observed, as a function of lattice size. 
\item
Extend our studies to SU(3) gauge theory using the techniques of \cite{Klco:2019evd}.
\item
Include dynamical quarks.
\end{itemize}
Obviously, realizing this agenda will require years of dedicated work. Our present contribution only marks its beginning.

\begin{acknowledgements}
We thank N.~Klco for insightful discussions. B.M. acknowledges support from the U.S. Department of Energy Office of Science (Grant DE-FG02-05ER41367) and from Yale University during extended visits. The authors gratefully acknowledge the scientific support and HPC resources provided by the Erlangen National High Performance Computing Center (NHR@FAU) of the Friedrich-Alexander-Universität Erlangen-Nürnberg (FAU) under the NHR project ID b172da. NHR funding is provided by federal and Bavarian state authorities. NHR@FAU hardware is partially funded by the German Research Foundation (DFG) – 440719683. X.Y. was supported in part by the U.S. Department of Energy, Office of Science, Office of Nuclear Physics, InQubator for Quantum Simulation (IQuS) (https:// iqus.uw.edu) under Award Number DOE (NP) Award DE-SC0020970 via the program on Quantum Horizons: QIS Research and Innovation for Nuclear Science. This work was partially facilitated through the use of advanced computational, storage, and networking infrastructure provided by the Hyak supercomputer system at the University of Washington.
\end{acknowledgements}

\appendix
\section{Energy level density on plaquette chains with $\jmax=\frac{1}{2}$}
\label{app:g_dependence}

\begin{figure}[ht]
\subfloat[\label{fig:spectrum_2d_13}]{%
  \includegraphics[height=1.25in]{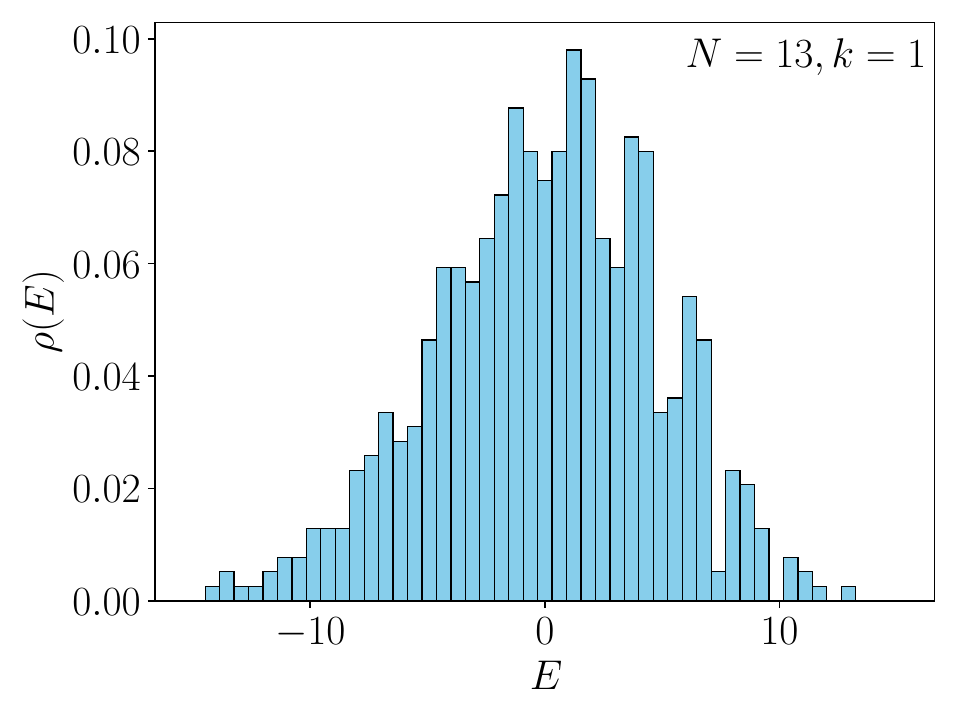}%
}
\hfill
\subfloat[\label{fig:spectrum_2d_19}]{%
  \includegraphics[height=1.25in]{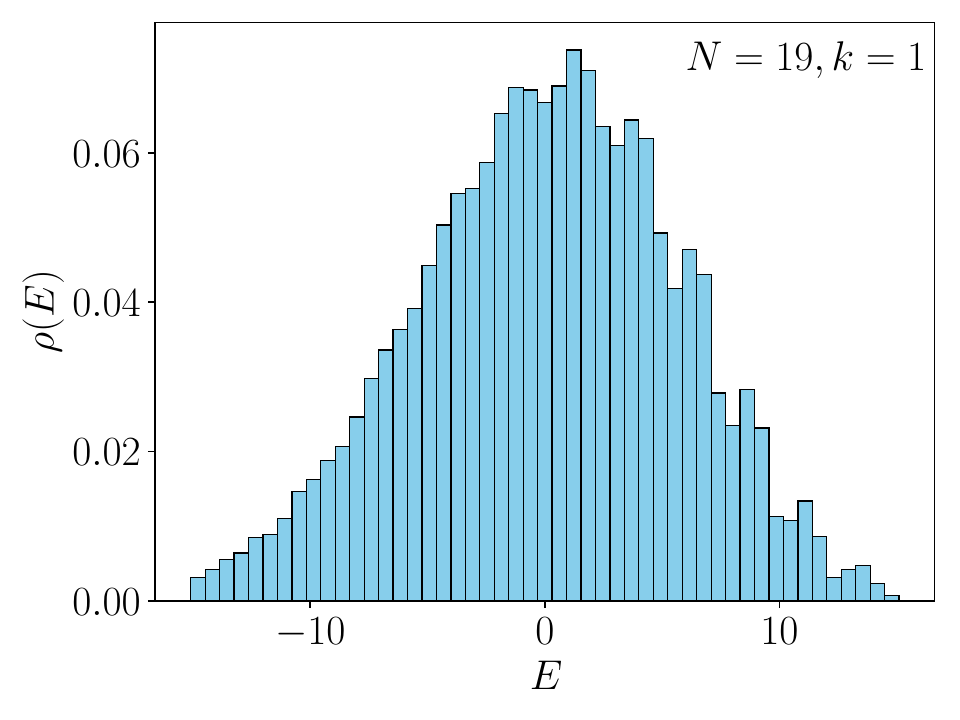}%
}
\caption{Normalized energy eigenvalue density $\rho(E)$ in the $k=1$ sector on the plaquette chain with $\jmax=\frac{1}{2}$ and $g^2=1$ for two different lattice sizes: (a) $N=13$, (b) $N=19$.}
\label{fig:eigenspectrum_N}
\end{figure}

In Fig.~\ref{fig:eigenspectrum_N}, we plot the normalized energy level density $\rho(E)$ in the $k=1$ sector on the plaquette chain with $\jmax=\frac{1}{2}$ and $g^2=1$ for two lattice sizes $N=13,19$. As can be seen, the level density becomes smoother as the lattice size increases, such that the behavior of the system should agree better with ETH predictions. However, the level density depends also on $g^2$ as can be seen by comparing the $N=19$ case for $g^2=1$ (Fig.~\ref{fig:spectrum_2d_19}) with that of Fig.~\ref{fig:EWindows} for $g^2=1.2$. Obviously, the spectrum is smoother for the latter case, indicating that for the former case $N$ has to be chosen larger to get the same quality results.

\section{Thermal properties}

\begin{figure}[ht]
	\centering
\includegraphics[width=0.95\linewidth]{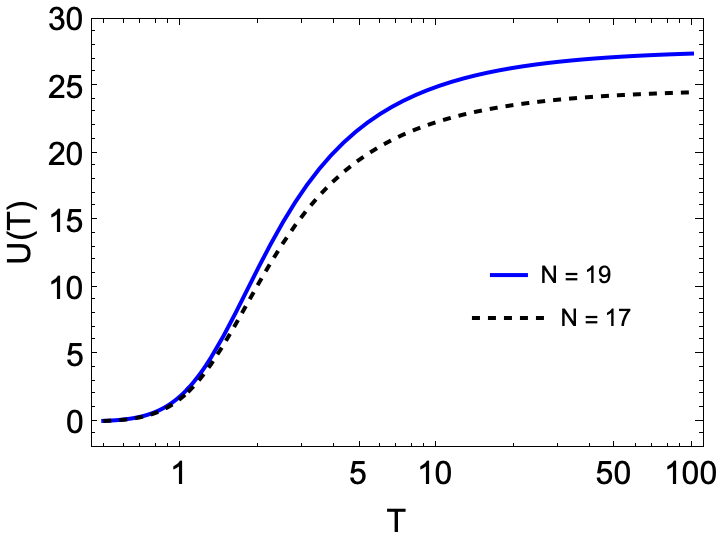}
	\caption{Internal energy $U({\bf \rm T})$ as a function of temperature ${\bf \rm T}$ for $N=19$ (solid, blue) and $N=17$ (dashed, black) plaquette chains with $\jmax\ = \frac{1}{2}$. $U({\bf \rm T})$ saturates at large ${\bf \rm T}$ because the system contains only a finite number of degrees of freedom. All quantities are shown in lattice units.}
	\label{fig:Thermo}
\end{figure}

It is instructive to study the thermal properties of some of our truncated lattice models. Figure~\ref{fig:Thermo} depicts the internal energy $U$ of the linear plaquette chain with $\jmax=\frac{1}{2}$ for $N=19$ (solid blue line) and $N=17$ (black dashed curve) as functions of the temperature ${\bf \rm T}$ in lattice units. The flattening out of $U({\bf \rm T})$ at high temperatures is an artifact of the finite lattice spacing, which eliminates high-momentum modes of the gauge field. This is even more apparent in the entropy $S({\bf \rm T})$, which approaches the value $S_\infty = N\ln 2$ for ${\bf \rm T}\to\infty$ as all energy levels are democratically populated. In the continuum, for the one-dimensional plaquette chain and without electric field cutoff, $U({\bf \rm T})$ would grow quadratically with ${\bf \rm T}$ at large ${\bf \rm T}$. The figure clearly shows that the system is still far from the continuum limit.

\begin{figure}[ht]
	\centering
 	\includegraphics[width=0.95\linewidth]{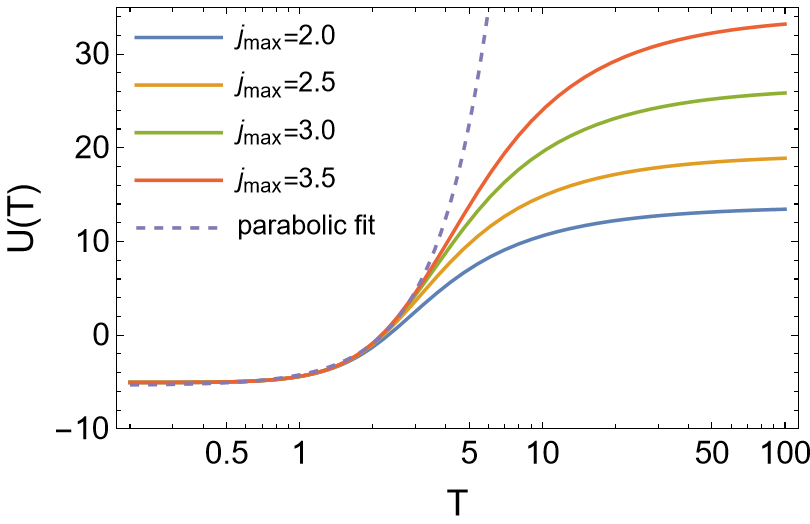}
	\caption{Internal energy $U({\bf \rm T})$ for the three-plaquette chain and increasing $\jmax$ at $g^2=1$. The quantity saturates at large ${\bf \rm T}$ because the system contains only a finite number of degrees of freedom.}
	\label{fig:Thermo_cutoff}
\end{figure}

Similarly we investigate the internal energy $U({\bf \rm T})$ for the $N=3$ plaquette chain with different momentum cutoffs $\jmax\in\{2, \frac{5}{2}, 3, \frac{7}{2}\}$ in Fig.~\ref{fig:Thermo_cutoff}. Obviously, the quadratic temperature dependence of $U({\bf \rm T})$ in the continuum theory is only attained in the $\jmax\rightarrow \infty$ limit, illustrating the necessity to perform this limit (as well as the $N\rightarrow \infty$ limit).

\bibliographystyle{apsrev4-1}
\bibliography{ETHforSU2}
\end{document}